%% file: main.tex
\documentclass[sigplan, nonacm, screen]{acmart}

\input{config}

\newcommand{\sys}{\texttt{UltraEP}\xspace}

\begin{document}

\title{\sys: Unleash MoE Training and Inference on Rack-Scale Nodes with Near-Optimal Load Balancing}

\newcommand{\authornotetext}[2]{%
  \begingroup
  \renewcommand{\thefootnote}{#1}%
  \footnotetext{#2}%
  \endgroup
}

\makeatletter
\g@addto@macro\@authornotes{%
  \authornotetext{$*$}{Work done during internship at Xiaohongshu Inc.}%
  \authornotetext{\textdagger}{Corresponding authors.}%
}
\makeatother

\newcommand{\pkuinstitution}{School of Computer Science, Peking University}

\author{
    \rm{
        Xinming Wei\textsuperscript{1}$^{*}$ \enskip
        Chao Jin\textsuperscript{1} \enskip
        Tuo Dai\textsuperscript{2} \enskip
        Yinmin Zhong\textsuperscript{1} \enskip
        Shan Yu\textsuperscript{3} \enskip
        Chengxu Yang\textsuperscript{4} \enskip
        Bingyang Wu\textsuperscript{1} \enskip
        \\
        Zili Zhang\textsuperscript{1} \enskip
        Jing Mai\textsuperscript{1} \enskip
        Qianchao Zhu\textsuperscript{4} \enskip
        Zhouyang Li\textsuperscript{4} \enskip
        Yuliang Liu\textsuperscript{4}\textsuperscript{\textdagger} \enskip
        Guojie Luo\textsuperscript{1}\textsuperscript{\textdagger} \enskip
    }
    \\
    \vspace{0.06in}
    {\textit{\textsuperscript{1}\pkuinstitution \quad
    \textsuperscript{2}Xiaohongshu Inc.}}
    \\
    {\textit{\textsuperscript{3}Shanghai AI Laboratory \quad
    \textsuperscript{4}Independent Researcher}}
}

\makeatletter
\gdef\authors{Xinming Wei, Chao Jin, Tuo Dai, Yinmin Zhong, Shan Yu, Chengxu Yang, Bingyang Wu, Zili Zhang, Jing Mai, Qianchao Zhu, Zhouyang Li, Yuliang Liu, and Guojie Luo}
\makeatother

\renewcommand{\shortauthors}{Wei et al.}

\input{sections/0-abstract.tex}

\maketitle

\input{sections/1-introduction.tex}
\input{sections/2-background.tex}
\input{sections/3-motivation.tex}
\input{sections/4-method-impl.tex}
\input{sections/5-evaluation.tex}
\input{sections/6-epilogue.tex}


\newpage
\bibliographystyle{ACM-Reference-Format}
\bibliography{main}


\end{document}

%% file: config.tex
\newcommand{\secref}[1]{\mbox{\S\hspace{0.1em}\ref{#1}}}

\newcommand{\figref}[1]{Fig.~\ref{#1}}

\newcommand{\tabref}[1]{Table~\ref{#1}}

\newcommand{\algoref}[1]{Algorithm~\ref{#1}}

\definecolor{textred}{RGB}{215,25,28}
\definecolor{textgreen}{RGB}{26,150,65}

\usepackage[ruled,vlined,linesnumbered]{algorithm2e}

\usepackage[T1]{fontenc}
\usepackage{xspace}

\usepackage{booktabs}
\usepackage{makecell}
\usepackage{threeparttable}
\usepackage{listings}
\usepackage{xcolor}
\AtBeginDocument{%
  \hypersetup{%
    linkcolor=ACMDarkBlue,
  }%
}
\usepackage{enumitem}
\usepackage{soul}
\usepackage[normalem]{ulem}
\usepackage{mathtools}

\usepackage[most]{tcolorbox}

\definecolor{codegray}{gray}{0.95}
\definecolor{commentgreen}{rgb}{0,0.5,0}
\definecolor{keywordblue}{rgb}{0.0,0.0,0.7}

\newcommand{\paraf}[1]{\noindent\textbf{#1}}
\newcommand{\parabf}[1]{\smallskip\noindent\textbf{#1}}

%% file: sections/0-abstract.tex
\begin{abstract}

Large-scale expert parallelism (EP) is becoming pivotal for training and serving frontier MoE models, but it also amplifies device-level expert load imbalance into compute stragglers, token all-to-all bottlenecks, and activation-memory spikes.
Existing balancers redistribute experts periodically based on historical load, which becomes unreliable for production deployments with non-stationary load patterns.

We present \sys, the first exact-load, real-time balancer for large-EP MoE training and serving prefill on rack-scale nodes (RSNs).
Leveraging the extended scale-up connectivity among dozens of GPUs within RSNs, \sys rebalances every microbatch and layer on critical paths, which requires nontrivial co-design of plan solving and expert replication communication to minimize exposed overhead.
To this end, \sys eagerly reacts to post-gating load with an efficient quota-driven planner, and executes the resulting irregular expert-state transfers with RSN-native persistent tile streaming and relay-based fan-out mitigation.
We evaluate \sys in a multi-RSN deployment of up to 256 GPUs, using cutting-edge MoE models from 106B to 671B parameters.
Averaged across training and serving, \sys achieves 94.3\,\% of the force-balanced ideal throughput, delivering 1.49$\times$ improvement over no-balancing, while reducing the final inter-rank imbalance from 1.30--4.01 to 1.01--1.04.
\end{abstract}

%% file: sections/1-introduction.tex
\section{Introduction}

Mixture-of-Experts (MoE) models have become a dominant paradigm for scaling large language models (LLMs), offering high model capacity with sparse activation~\cite{shazeer2017outrageously, lepikhin2020gshard, fedus2022switch, dai2024deepseekmoe, deepseek2024v3, yang2024qwen2, yang2025qwen3}.
To accommodate larger MoE models, expert parallelism (EP) is widely adopted, where experts are distributed across devices and tokens are dynamically routed via all-to-all communication~\cite{lepikhin2020gshard, rajbhandari2022deepspeed, hwang2023tutel, gale2023megablocks}.
Compared to other forms of model parallelism, EP scales more efficiently, exposing abundant parallelism while maintaining high arithmetic intensity.
In production, large-scale expert parallelism (large-EP), such as 32- or 64-way EP, has been indispensable for training and serving MoE models with hundreds of billions of parameters~\cite{mcore2026, deepseek2024v3, sglang-large-ep-h100, vllm-wide-ep-blackwell}.

However, large-EP amplifies a fundamental challenge: expert load imbalance.
As illustrated in \figref{fig:bg-ep}, tokens are unevenly distributed across experts due to input diversity and routing dynamics, leading to skewed workload across devices~\cite{wang2024auxiliary,nie2023flexmoe,zeng2025efficientmoe}.
This imbalance manifests in expert computation stragglers, token all-to-all bottlenecks, and activation memory spikes on overloaded devices.
As the EP degree increases, these effects compound, significantly widening the gap between achieved and ideal performance~\cite{mcore2026}.

\begin{figure}[t]
  \centering
  \includegraphics[width=0.96\linewidth]{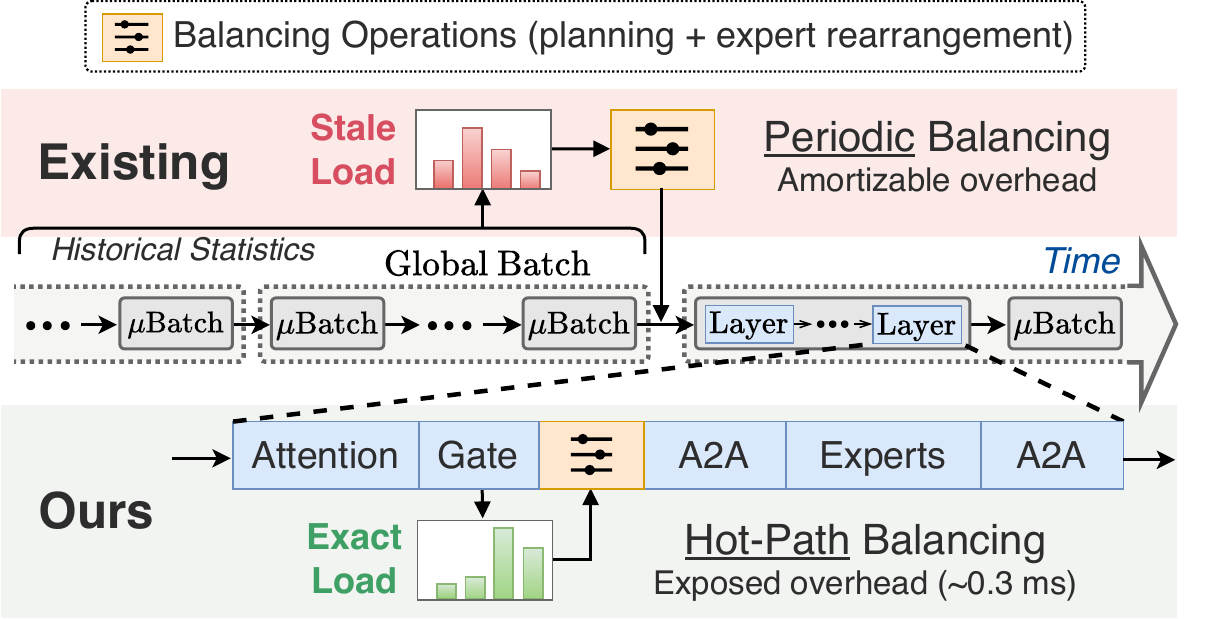}
  \caption{
    \sys differs from prior solutions in load fidelity, decision timing, and balancing frequency.
  }
  \Description{Overview comparing UltraEP with prior balancing approaches across load fidelity, decision timing, and balancing frequency.}
  \label{fig:teaser}
\end{figure}

Existing solutions mainly mitigate imbalance by predicting incoming load and adjusting expert-to-device placement accordingly~\cite{zhang2025popfetcher, eplb,yang2026libra,zhai2023smartmoe}.
EPLB~\cite{eplb}, a widely used balancer, adopts a redundant expert strategy, heuristically replicating high-load experts on multiple devices.
Although EPLB is agnostic to load estimator, common deployments~\cite{eplb-vllm, eplb-sgl} use recent routing history and rebalance periodically to amortize the planning and rearrangement overhead (\figref{fig:teaser}).
Its effectiveness, however, depends on the stationarity of the load patterns.
On frontier MoE models with hundreds of experts, we observe that expert popularity shifts sharply across microbatches, layers, and data domains in both training and serving (\secref{sec:load-profile}).
Stale predictions then produce inaccurate placements, leaving substantial residual imbalance (\figref{fig:profile-acc}).

This observation motivates a shift from prediction-based to exact-load balancing (\figref{fig:teaser}).
However, exact load becomes available only after gating, forcing balancing operations onto the critical path.
The exposed overhead includes online plan solving and heavy expert rearrangement communication, with weight transfers in forward execution and additional gradient or optimizer-state movement in training.
In standard RDMA clusters, high-bandwidth scale-up connectivity is confined to a single 4/8-GPU node, while inter-node traffic relies on slower scale-out networks.
Under large-EP, moving substantial expert states across multiple nodes is prohibitively expensive and impractical on the hot path.

The emergence of rack-scale nodes (RSNs)~\cite{nvidia-blackwell, nvidia-rubin, amd-helios, cm384-serving} fundamentally changes this design space.
By extending the scale-up links across dozens of GPUs in a full rack, RSNs can keep an entire EP group on a high-bandwidth domain, making hot-path balancing physically viable.
\figref{fig:bg-rsn} depicts this paradigm shift.
However, RSNs are necessary but not sufficient, with two pivotal \uline{challenges}:
The \emph{control plane} must make a high-quality balancing decision within the short window between gating and token dispatch.
The \emph{data plane} must then execute irregular, volatile expert state transfers that are poorly backed by static collectives, potentially underutilizing RSN bandwidth.
Without careful co-design, these overheads can easily negate the balancing gain.

We present \sys, the first system to achieve \emph{exact} expert load balancing in real-time for large-EP MoE training and serving prefill on RSNs, attaining near force-balanced ideal throughput.
As shown in \figref{fig:teaser}, it rebalances eagerly at the granularity of each microbatch and layer with minimal hot-path latency.
It is GPU-native without host-side bottlenecks, and optimized for the unique communication patterns of expert replication on RSNs.
With dedicated memory layout (e.g., reusing redundant expert buffers across layers), it reduces memory overhead by dozens of times.
It preserves training equivalence by reducing replica gradients back to main experts before optimizer updates.

\sys comprises two key \uline{innovations} to tackle the control- and data-plane challenges.
First, we propose a quota-driven planner that jointly optimizes expert replication and token reroute (which redirects each token of a replicated expert to one of its physical instances).
Each quota specifies the final token load assigned to an expert instance.
These quotas couple replica creation and token reroute: a replica is materialized only when it carries useful load, and reroute then realizes the solved quota split with locality.
Unlike EPLB, which rebalances on stale load and leaves reroute to a separate heuristic (e.g., round-robin), \sys directly optimizes the post-reroute load with an efficient threshold-based binary search.
Second, \sys tailors RSN-native communication for dynamic expert traffic induced by real-time replication.
It streams expert transfers as device-side tile-level tasks to saturate the bandwidth.
For hot experts with many replicas, \sys builds chunk-streaming relay trees that split hotspot fan-out traffic across ranks with spare bandwidth.

\sys is designed for \emph{production} deployment on multiple RSNs.
It is compatible with common tensor, pipeline, and data parallelism.
We implement \sys as a standalone runtime, allowing seamless integration into established training and serving stacks.
On 106B--671B MoE models and up to 256 GPUs, \sys sustains 94.6\,\% of the force-balanced ideal throughput in training and 93.9\,\% in serving prefill on average.
It also improves training throughput by an average of 1.42$\times$ over Megatron-LM~\cite{Shoeybi2019Megatron} and serving prefill throughput by 1.56$\times$ over SGLang~\cite{zheng2024sglang}, while keeping post-balancing inter-rank imbalance around 1.01--1.04.
\sys outperforms prevalent balancers, even strengthened EPLB with exact load, validating quota-driven planning.
It also accelerates expert replication by 3.1$\times$--5.5$\times$ over mainstream communication backends, showing the advantage of RSN-tailored communication.
In real-world production MoE training, \sys shows stable performance with over 92\,\% of ideal throughput while preserving convergence.

This paper makes the following contributions.
{
\setlength{\leftmargini}{1.2em}
\begin{itemize}
    \item We characterize non-stationary expert load imbalance in MoE training and serving prefill at scale (\secref{sec:load-profile}).
    \item We design \sys, the first exact-load balancer in real time for large-EP MoE deployment on RSNs (\secref{sec:method-overview}).
    \item We build quota-based planning (\secref{sec:method-solver}) and RSN-native expert-state communication (\secref{sec:method-comm}) for hot-path balancing.
    \item We validate \sys's near-optimal balancing quality, near-ideal throughput, and production scalability (\secref{sec:evaluation}).
\end{itemize}
}

%% file: sections/2-background.tex
\section{Background}

\subsection{Rack-Scale Node}

\begin{figure}[t]
    \centering
    \includegraphics[width=0.90\linewidth]{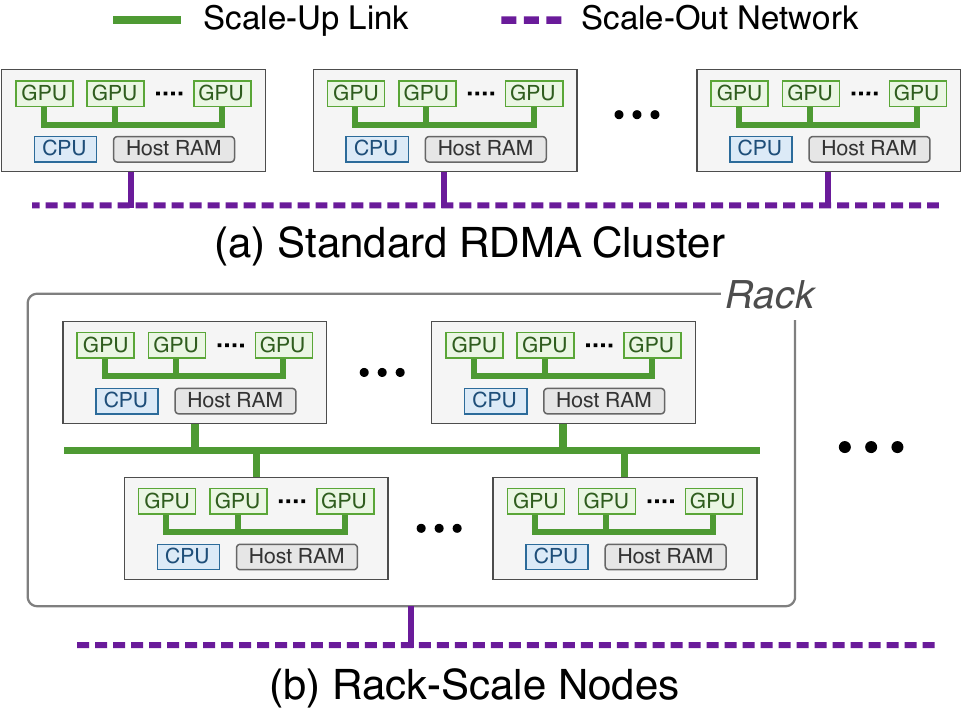}
    \caption{
        Illustration of expanded scale-up domain within a rack-scale node, compared with the standard RDMA cluster.
    }
    \Description{Comparison between a standard RDMA cluster and a rack-scale node with an expanded scale-up domain.}
    \label{fig:bg-rsn}
\end{figure}

As shown in \figref{fig:bg-rsn}, a rack-scale node (RSN) expands the scale-up domain from a single 4/8-GPU server to a full rack, typically spanning 64$+$ GPUs~\cite{nvidia-blackwell, nvidia-rubin, amd-helios, cm384-serving}.
An RSN is still composed of multiple servers, but GPUs across servers remain directly connected via a rack-wide scale-up fabric~\cite{nvidia-nvlink, ualink, ub-mesh}.
Compared with scale-out, scale-up offers much higher per-GPU bandwidth (hundreds of GB/s) and load/store-style memory semantics, whereas scale-out uses packet-based networking and typically provides only tens of GB/s per NIC.
For MoE models, an EP group can often be contained within one RSN, keeping expert dispatch on the fast scale-up fabric rather than the slower scale-out network.

\subsection{Distributed MoE Training and Inference}

\paraf{MoE Architecture Evolution.}
Early MoE models (e.g., GShard~\cite{lepikhin2020gshard}, Mixtral~\cite{jiang2024mixtral}, Switch Transformer~\cite{fedus2022switch}) adopt a coarse-grained design with a small number of large experts.
Each expert is a wide feed-forward network, and the gating network produces expert scores, selects a small top-$k$ subset of experts for each token, and uses the corresponding top-$k$ weights to scale and combine expert outputs.

Currently, frontier MoE models, including Google's Gemini-3 Pro~\cite{gemini3promodelcard}, OpenAI's GPT-OSS~\cite{openai2025gptoss}, DeepSeek's DeepSeekMoE~\cite{dai2024deepseekmoe} and DeepSeek-V3/R1~\cite{deepseek2024v3, deepseek2025r1}, Qwen MoE families~\cite{yang2024qwen2, yang2025qwen3}, and Meta's Llama-4~\cite{llama4modelcard}, have evolved toward fine-grained MoE models, utilizing hundreds of experts where each individual expert is smaller and computationally lighter.
This trend improves expert specialization and scaling flexibility, but it also makes routing distribution and expert load balancing much more dynamic.

\parabf{LLM Inference.}
LLM inference consists of two phases: \emph{prefill} and \emph{decode}.
Prefill is compute-bound, processing the prompt in parallel and filling the KV cache.
Decode is memory-bound because each step must repeatedly fetch large model weights and KV cache entries from memory, while the per-token computation is too small to hide these transfers.
These phases directly shape user-visible latency metrics such as time to first token (TTFT) and time per output token (TPOT), where TTFT is largely determined by prefill and TPOT reflects steady-state decode speed.
Established improvements include prefill-decode (PD) disaggregation~\cite{zhong2024distserve, patel2024splitwise} or chunked prefill~\cite{agrawal2024sarathi} to reduce PD interference.
PD disaggregation isolates the two phases on different resources or execution paths, while chunked prefill breaks long prompts into smaller pieces and batches them together with decode requests.

\begin{figure}[t]
    \centering
    \includegraphics[width=\linewidth]{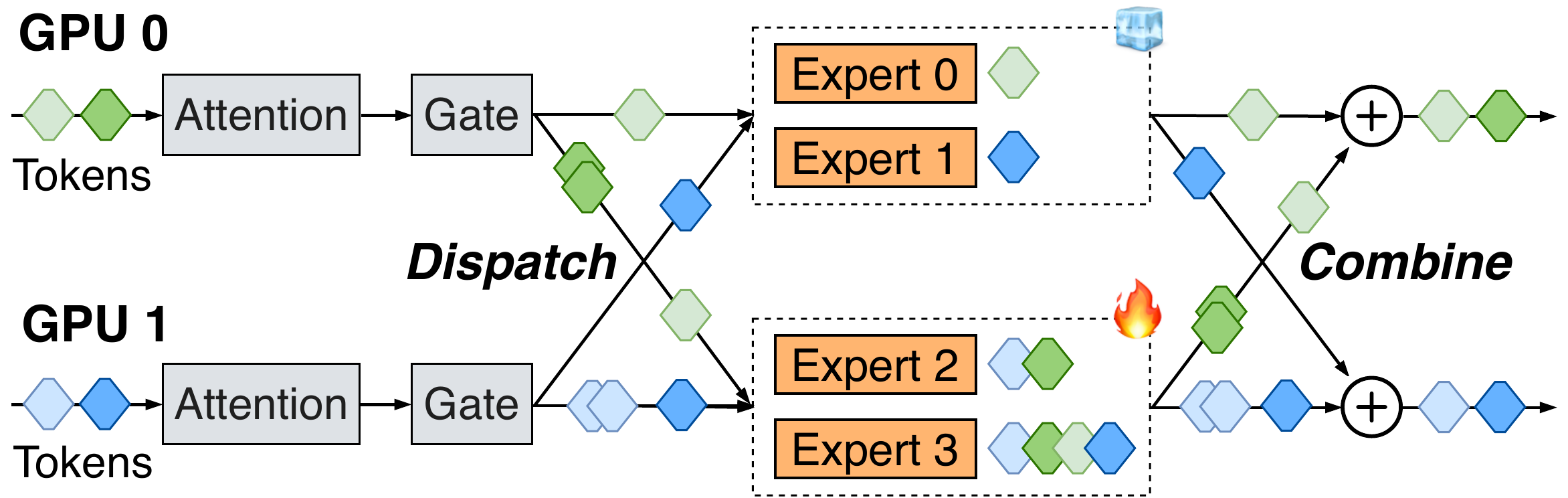}
    \caption{
        An illustrative example of MoE forward under expert parallelism (EP): 4 experts, $\text{EP}=2$, and $\text{top-}k=2$.
    }
    \Description{Example token routing and all-to-all communication for MoE forward execution under expert parallelism.}
    \label{fig:bg-ep}
\end{figure}

\parabf{Expert Parallelism (EP)}.
As illustrated in \figref{fig:bg-ep}, EP~\cite{lepikhin2020gshard} partitions experts across GPUs and routes tokens to the GPUs that host their selected experts.
Each rank first computes gating decisions for its local tokens, performs a global exchange of routing metadata to determine per-peer send/receive sizes and offsets, and then \emph{dispatches} tokens via all-to-all communication.
Upon receiving tokens, each rank groups them by expert, executes \emph{grouped GEMM} over the resulting expert batches, and returns the tokens through a \emph{combine} phase that mirrors dispatch.
Compared with general-purpose communication libraries, DeepEP~\cite{deepep} optimizes token all-to-all with high-performance kernels, topology-aware scheduling, and potential deduplication or overlapping.

MoE training and inference typically combine tensor parallelism (TP)~\cite{Shoeybi2019Megatron}, pipeline parallelism (PP)~\cite{huang2019gpipe, narayanan2019pipedream, li2021chimera, qi2024zerobubble}, and data parallelism (DP)~\cite{shallue2019measuring, rajbhandari2020zero}, with EP as the primary scaling axis for experts.
EP improves (1) computational efficiency by aggregating tokens to increase per-expert GEMM batch size, and (2) communication efficiency by keeping all-to-all volume independent of expert count.
In practice, EP is usually an inner parallel mode nested within an outer PP/DP layout, where each MoE layer forms an EP group.
Within each EP group, attention blocks are often replicated (attention-side DP~\cite{liu2025moe, jin2025megascale}), while experts are partitioned across GPUs and participate in the all-to-all token exchange.

\parabf{Relationship between Algorithm- and System-Side Load Balancing in MoE Training.}
Algorithm-side routing regularization and system-side balancing are complementary rather than interchangeable.
Training-time auxiliary routing losses primarily stabilize optimization, prevent \emph{routing collapse}~\cite{shazeer2017outrageously}, and preserve expert specialization.
For example, GShard encourages balanced utilization in expectation via an auxiliary loss that aligns each expert's average router probability with its realized token fraction~\cite{lepikhin2020gshard}, while DeepSeekMoE updates an expert-wise routing bias from recent load to downweight overloaded experts and upweight underutilized ones without interference gradients~\cite{deepseek2024v3,wang2024auxiliary}.
Both promote balance over time but cannot ensure per-microbatch realized load balance, especially for fine-grained MoE models under large-EP.
System-side techniques therefore correct runtime imbalance; they cannot replace routing losses because they do not serve the same modeling objectives.

%% file: sections/3-motivation.tex
\section{Expert Load Analysis}
\label{sec:load-profile}

We observe highly \emph{skewed}, \emph{heterogeneous}, and \emph{dynamic} expert load distributions in both serving and training.
This section illustrates these load patterns and evaluates the effectiveness of EPLB under such load and large-EP.

\begin{figure}[t]
    \centering
    \includegraphics[width=\linewidth]{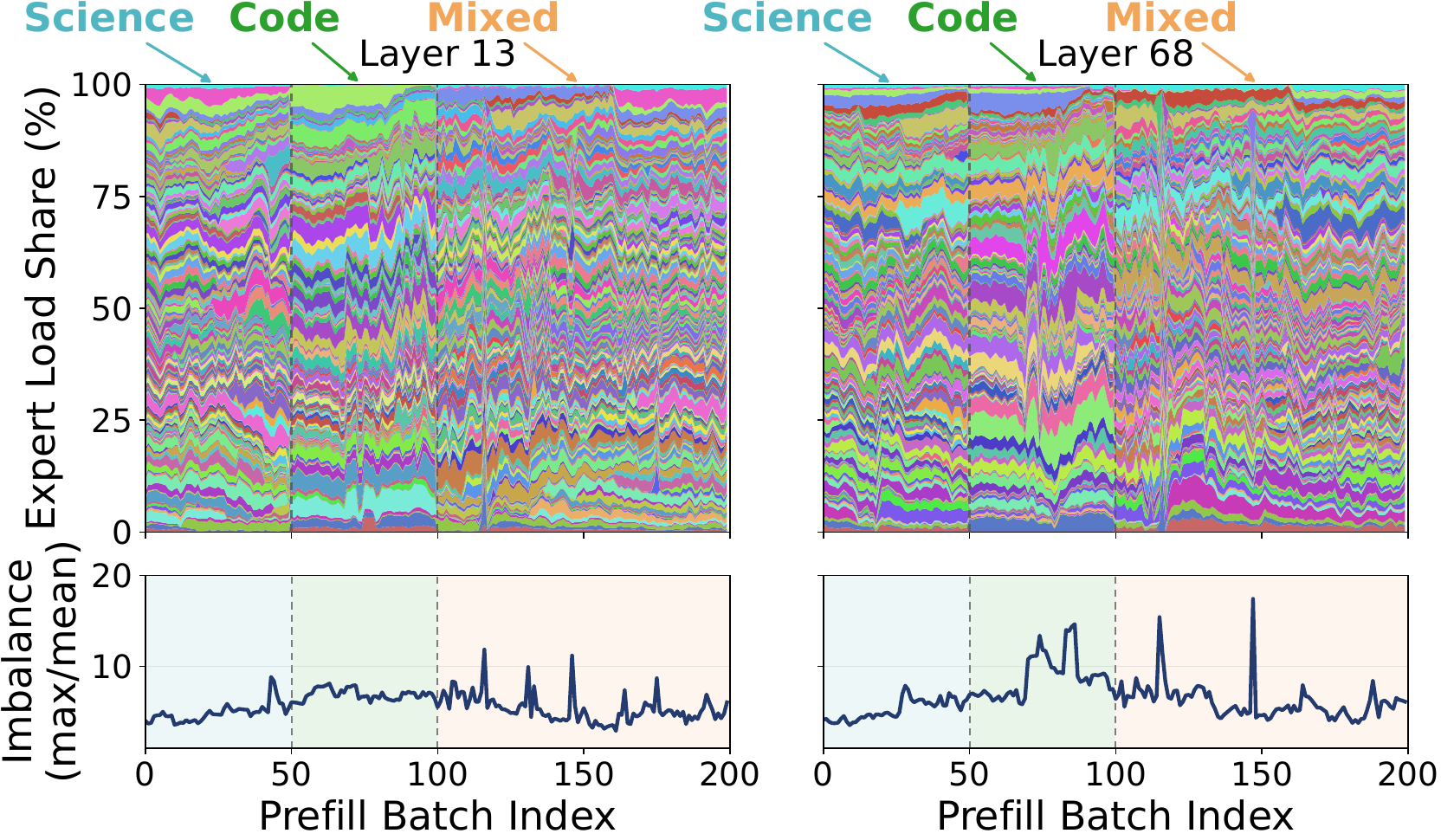}
    \caption{Prefill-time expert load distribution shifts across forward steps, data domains, and layers. Sampled on Qwen3-235B~\cite{yang2025qwen3} (top-8 activated of 128 experts) with EP=64. The imbalance ratio denotes the max per-\emph{expert} load by mean.}
    \Description{Prefill expert-load distributions showing shifts across steps, domains, and layers.}
    \label{fig:profile-serving}
\end{figure}

\parabf{Serving Prefill~(\figref{fig:profile-serving}).}
In serving, the prefill stage is the primary source of expert load imbalance.
For memory-bound decode, we discover that the impact of compute-side imbalance is largely diluted by memory access latency.
Increasing the batch size can improve compute intensity, but that conflicts with strict SLOs for decode TPOT~\cite{zhong2024distserve}.
Therefore, the practical balancing target is prefill, where throughput is prioritized to reduce TTFT.
As shown in \figref{fig:profile-serving}, expert popularity varies sharply across semantic transitions, including science, coding, and mixed-domain traffic.
Even within a single domain, the hot experts drift from one batch to the next, while mixed-domain inputs superimpose multiple routing patterns and make the imbalance even less predictable.
This yields a workload that is both highly skewed and non-stationary.

\begin{figure}[t]
    \centering
    \includegraphics[width=\linewidth]{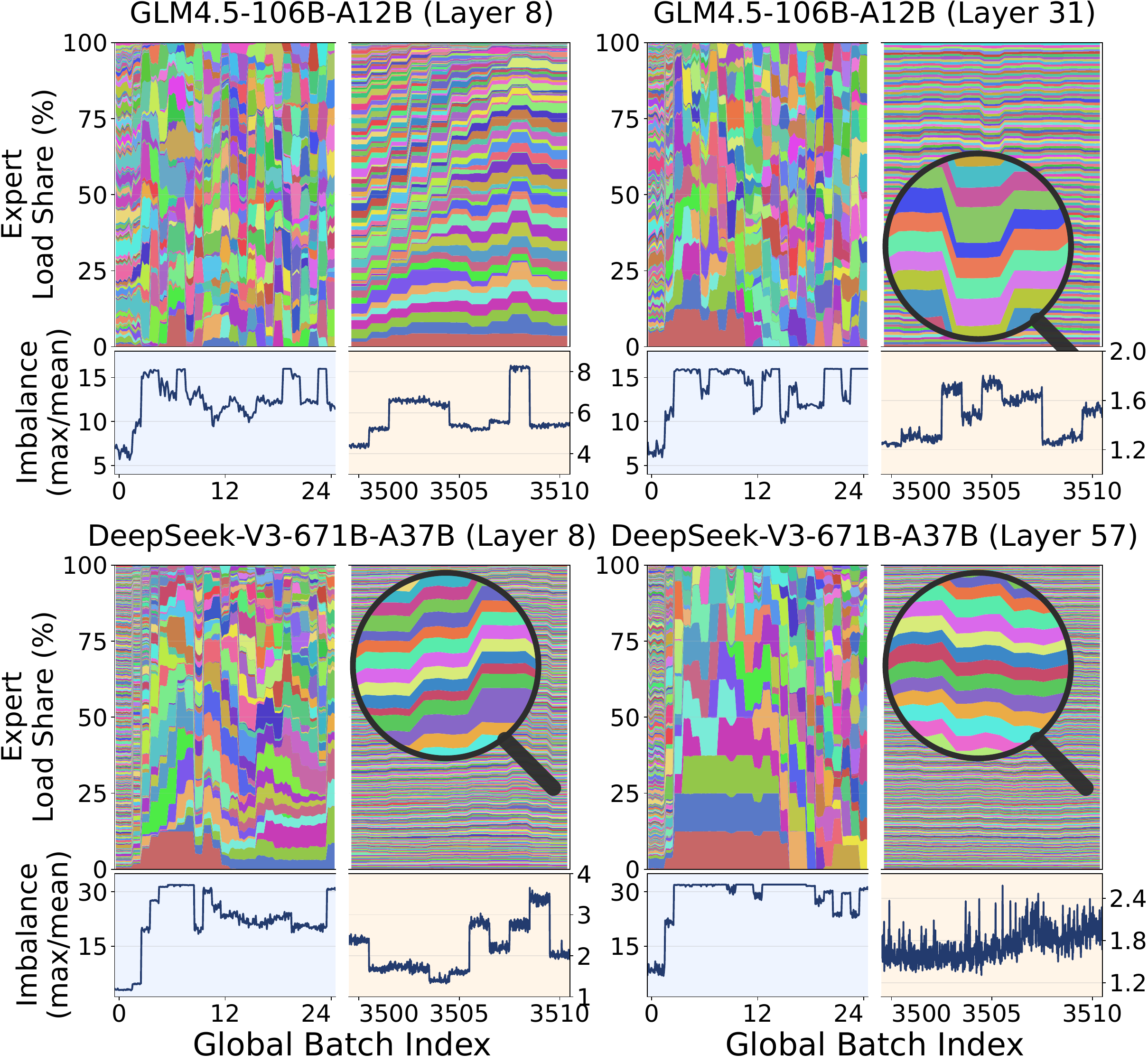}
    \caption{Training-time expert load distributions in the initial (first 25) and late (3500--3510 of 4500 total global batches) stages. Sampled on GLM4.5-106B-A12B~\cite{glm2025v45} (top-8 activated of 128 experts, trained with GShard-style auxiliary loss) and DeepSeek-V3~\cite{deepseek2024v3} (top-8 activated of 256 experts, using DeepSeek-style auxiliary loss) within one EP64 group.}
    \Description{Training expert-load distributions during early and late training stages for two MoE models.}
    \label{fig:profile-train}
\end{figure}

\parabf{Training~(\figref{fig:profile-train}).}
Training exhibits a different yet equally challenging pattern.
Early in training, expert loads are highly unstable because router specialization has not yet stabilized.
As training proceeds, the average distribution becomes smoother, but strong dynamics remain.
Across \emph{global batches}, auxiliary-loss negative feedback continually re-adjusts expert utilization; even DeepSeek-style router compensation that proactively equalizes experts does not eliminate the oscillation.
Inter-\emph{microbatch} jitter from sampling randomness also remains visible at a finer granularity, which is even more pronounced on DeepSeek-V3 training.

\begin{figure}[t]
    \centering
    \includegraphics[width=0.95\linewidth]{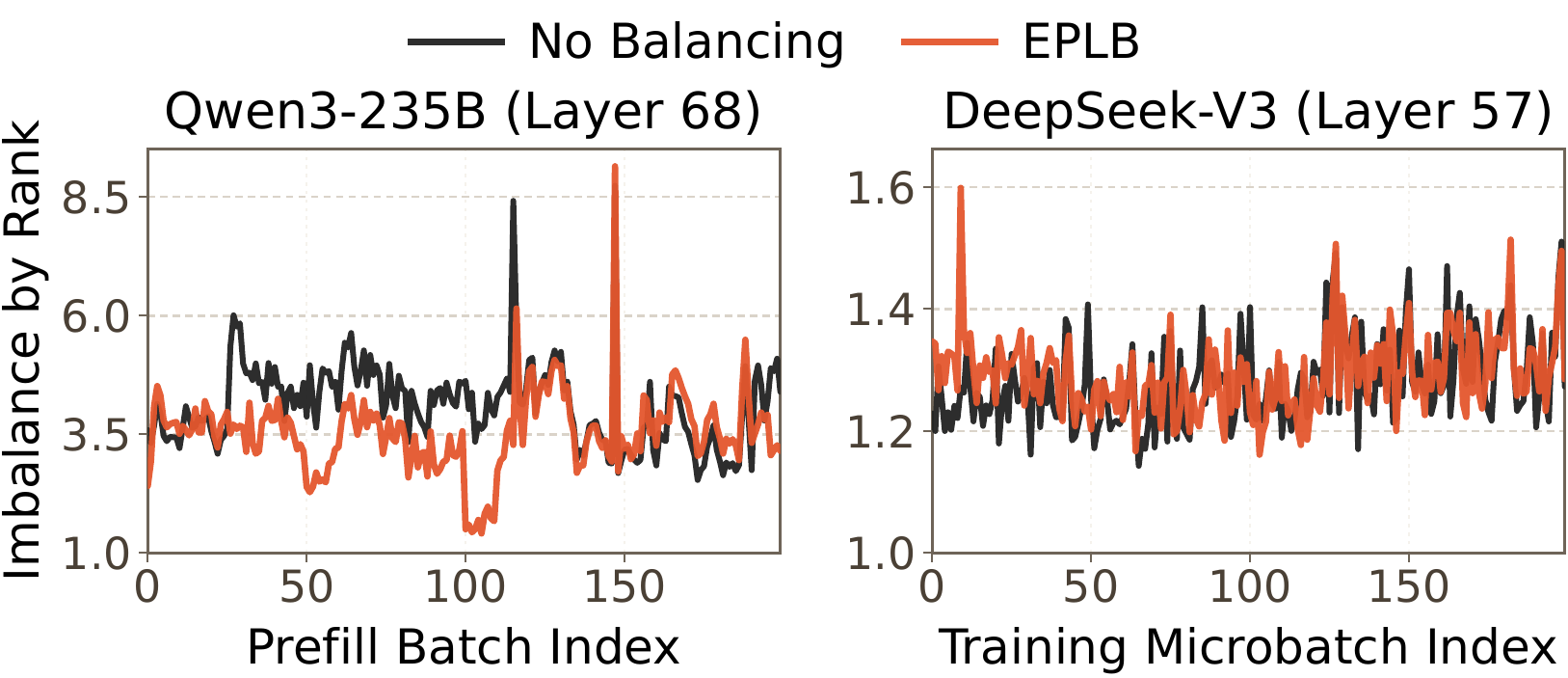}
    \caption{\emph{Rank}-level imbalance before and after EPLB, computed from previously recorded loads with EP=64.
    EPLB rebalancing interval is 50 batches for prefill and 3 global batches for training, respectively.
    Prefill (left) uses mixed data, while training (right) shows the 3510$^{\mathrm{th}}$ global batch.
    }
    \Description{Rank-level imbalance before and after EPLB for prefill and training workloads.}
    \label{fig:profile-acc}
\end{figure}

\parabf{Limitations of History-Based Balancing~(\figref{fig:profile-acc}).}
We study how EPLB mitigates rank-level imbalance under large-EP.
With fewer experts per EP rank, large-EP directly translates routing dynamics across experts into pronounced inter-rank skew.
However, EPLB relies on historical load statistics and cannot track fast, non-stationary load shifts in either prefill or training.
When the realized load deviates from the statistics used to derive the expert layout, EPLB can even worsen imbalance, creating spikes and new stragglers.
This motivates a real-time design that reacts to the instantaneous load rather than extrapolating from stale measurements.

%% file: sections/4-method-impl.tex
\section{\sys System Design}
\label{sec:method-overview}
\sys targets real-time expert load balancing with exact load.
Each EP group is inside an RSN scale-up domain, while cross-rack scale-out expansion uses PP and DP.
\sys is tailored for both training and serving prefill.
It solves expert replication and token rerouting plans on the fly, performs weight distribution to materialize redundant experts, and aggregates replicas' gradients during the corresponding backward passes in training.
This section first introduces \sys's expert placement and memory management~(\secref{subsec:overview-layout}), then overviews the forward/backward execution pipeline~(\secref{subsec:overview-pipelines}), formulates the per-layer real-time optimization problem~(\secref{subsec:overview-formulation}), and finally distills the practical control-plane and data-plane challenges~(\secref{subsec:overview-challenges}).

\begin{table}[t]
\centering
\caption{Notation used in \sys.}
\label{tab:sys-notation-yushan}
{\footnotesize
\setlength{\tabcolsep}{1pt}
\begin{tabular}{@{}p{0.20\linewidth}p{0.79\linewidth}@{}}
\toprule
\textbf{Symbol} & \textbf{Meaning} \\
\midrule
$\mathcal{R}$ & All ranks in one EP group, with size $R:=|\mathcal{R}|$. \\
$h(e)$ & Home rank of logical expert $e$, hosting its main instance. \\
$\mathcal{H}(e)$ & Host ranks of expert $e$'s physical instances. \\
$N_{\mathrm{layers}}$ & Number of MoE layers in the underlying model. \\
$N_{\mathrm{slot}}$ & Number of redundant slots on each rank. \\
$\mathcal{E}$ & All logical experts in one EP group. \\
$\mathcal{E}_r$ & Main experts on rank $r$, i.e., $\{e\in\mathcal{E}\mid h(e)=r\}$. \\
$\Lambda=\{\lambda_{r,e}\}$ & Global load matrix; $\lambda_{r,e}$ is the token load from source rank $r$ assigned to logical expert $e$. \\
$U=\{u_{e, r}\}$ & Solved load quota table; $u_{e,r}>0$ iff rank $r$ hosts a physical instance of expert $e$, which carries post-reroute load $u_{e,r}$. \\
$X=\{x_{r,s}\}$ & Redundant slot assignment; $s\in[N_{\mathrm{slot}}]$, $x_{r,s}=e$ if slot $s$ on rank $r$ hosts a replica of expert $e$, and otherwise $\varnothing$. \\
$Q=\{q_{r,e,t}\}$ & Reroute split from source rank $r$ to expert $e$'s physical instance on rank $t$, satisfying $\sum_{t\in\mathcal{H}(e)} q_{r,e,t}=\lambda_{r,e}$ and $\sum_{r\in\mathcal{R}} q_{r,e,t}=u_{e,t}$. \\
$u_{\min}$ & Minimum useful quota of a newly created replica (set to 1024 in this work). \\
$\beta$ & Target balancing coefficient (set to 1.01 in this work). \\
\bottomrule
\end{tabular}
}
\end{table}

\begin{figure}[t]
    \centering
    \includegraphics[width=\linewidth]{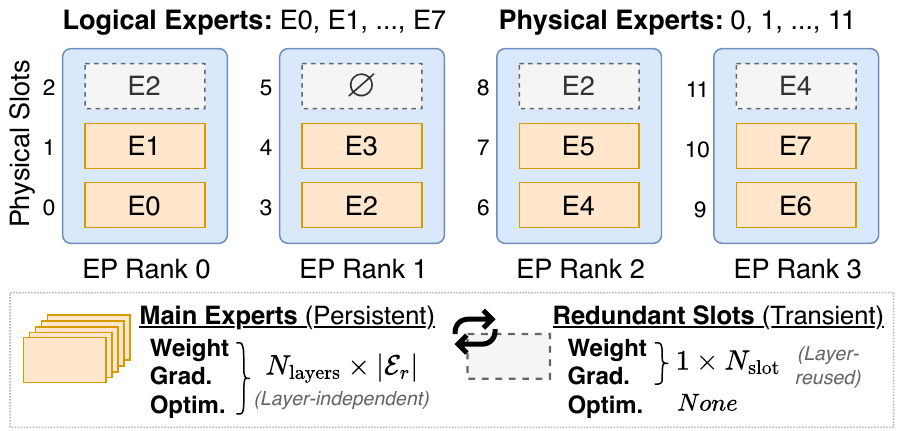}
    \caption{Expert layout and buffer management (example: 8 experts, single layer, $\text{EP}=4$, $N_{\text{slot}}=1$).
    Redundant expert slots reuse weight and gradient buffers across layers, with no optimizer state.
    Main experts retain the full set of buffers.
    }
    \Description{Expert layout showing main expert slots, redundant expert slots, and cross-layer buffer reuse.}
    \label{fig:overview-layout}
\end{figure}

\subsection{Expert Layout and Memory Management}\label{subsec:overview-layout}
As illustrated in \figref{fig:overview-layout}, we use \emph{logical expert} to denote the expert identity defined by the model, and \emph{physical expert} to denote an expert replica instantiated by a rank.
Every rank reserves the same number of \emph{main} and \emph{redundant} slots.
A main slot hosts the original physical instance of a logical expert, whereas a redundant slot either hosts one replica of a logical expert or remains empty.
This fixed layout keeps the runtime clean and deterministic, and yields a one-to-many logical-to-physical mapping: each logical expert has one fixed main instance and zero or more redundant replicas.

\parabf{Replication Only.}
\sys adopts replication-only balancing.
It never reorders main experts.
This reordering-free design is effective because large-EP reduces the number of local main experts per rank (often two or four).
At that point, reordering brings diminishing marginal benefits; it can only reshuffle a tiny local set while incurring substantial state migration, control complexity, and locality disruption.
In contrast, the cost-effective replication expands the service capacity of the actual bottlenecks.

\parabf{Cross-Layer Buffer Reuse.}
For main experts, \sys preserves the standard training/serving memory layout.
For each redundant slot, it keeps no optimizer state (optimizer updates are applied only on main experts) while sharing weight/gradient buffer across layers.
In Qwen3-235B-A22B (94 MoE layers, 128 experts), this reduces a single redundant slot from 3.3\,GB weights and 6.6\,GB gradients to 36\,MB and 72\,MB per rank, at the cost of a tight, per-layer weight-materialization deadline on the forward critical path~(\secref{subsec:overview-pipelines}).

\subsection{Computation-Communication Pipelines}\label{subsec:overview-pipelines}

\begin{figure}[t]
    \centering
    \includegraphics[width=\linewidth]{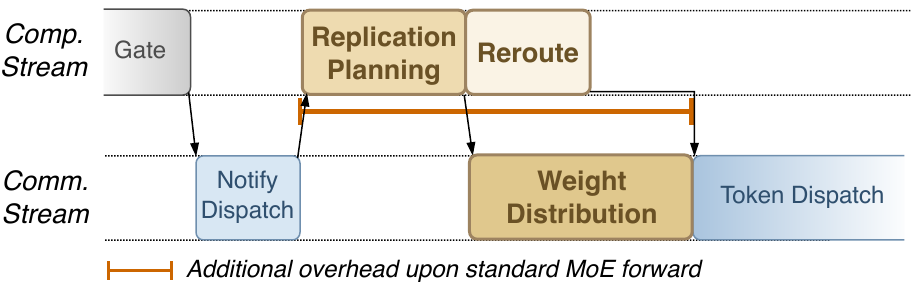}
    \caption{MoE forward pass with \sys enabled.}
    \Description{Timeline of the MoE forward pass with planning, weight distribution, reroute, token all-to-all, and expert computation.}
    \label{fig:overview-fwd}
\end{figure}

\paraf{Forward: Eager Planning and Expert Replication~(\figref{fig:overview-fwd}).}
\sys first reuses the existing notify-dispatch for later token all-to-all, to gather global routing information.
Given this exact load, every rank deterministically computes an identical replication and reroute plan with no extra synchronization.
Reroute converts the router output from token-to-logical-expert assignments into token-to-physical-expert.
Both are fully on-device without host bottlenecks.
Once replication is decided, \sys distributes main-expert weights to their remote replicas on each rank.
This synchronization can overlap with reroute, but token dispatch should wait for it to finish to avoid bandwidth contention.
As a result, planning and weight replication both stay on the critical path.
This imposes strict timeliness requirements.

\begin{figure}[t]
    \centering
    \includegraphics[width=\linewidth]{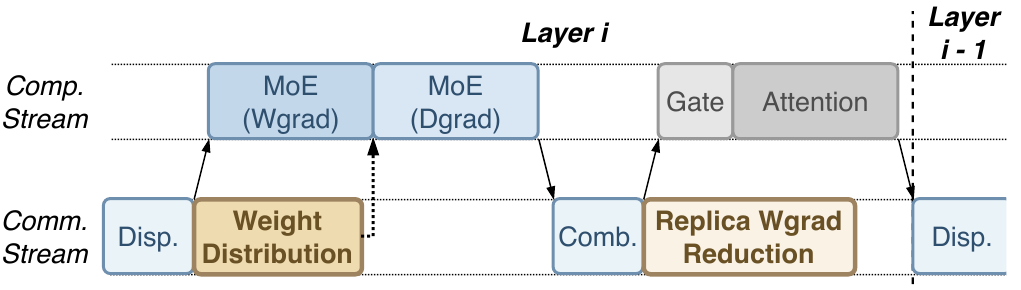}
    \caption{MoE backward pass with \sys enabled.}
    \Description{Timeline of the MoE backward pass with weight re-materialization, gradient computation, and gradient reduction.}
    \label{fig:overview-bwd}
\end{figure}

\parabf{Backward: Weight Re-materialization and Gradient Reduction~(\figref{fig:overview-bwd}).}
Backward execution starts by restoring the redundant expert weights to the same state in forward pass.
This communication can overlap with weight gradient (Wgrad) computation, until the start of data gradient (Dgrad) to avoid race conditions.
After MoE backward finishes, every main expert aggregates the gradients contributed by all of its remote replicas into the main gradient buffer.
This reduction preserves equivalence with the no-replica formulation and must finish before the next MoE layer begins, since the redundant gradient buffer is also reused across layers.
Backward execution does not solve replication again, and it reuses the cached metadata from the forward pass.
The reverse reroute is effectively a scatter-to-gather inversion of the forward assignment with negligible overhead.

\subsection{Problem Formulation}\label{subsec:overview-formulation}

We formulate an online optimization problem solved independently for each EP group. Table~\ref{tab:sys-notation-yushan} introduces the notation.

\parabf{Input.}
Static metadata $(\mathcal{R},\mathcal{E},h,N_{\mathrm{slot}})$ and runtime load $\Lambda$.

\parabf{Output.}
A quota-aware replication plan $U=\{u_{e,t}\}$ and a reroute split $q_{r,e,t}$ consistent with $U$.

\parabf{Objective.}
The forward objective is to minimize
\begin{equation}
T_{\text{solve\_rep}}^{fwd}+\max\!\left(T_{\text{reroute}}^{fwd},T_{w\_\text{distr}}^{fwd}\right)+T_{\text{tok\_a2a}}^{fwd}+T_{\text{moe}}^{fwd}.
\end{equation}
Here the terms denote the latencies of plan solving, reroute, weight distribution, token all-to-all, and MoE compute.

The corresponding backward objective is
\begin{equation}
\min \quad T_{\text{tok\_a2a}}^{bwd}+T_{\text{moe}}^{bwd}.
\end{equation}

Backward reuses forward replication and reroute metadata, with replica-specific communication hidden under computation. Thus, only token all-to-all and MoE compute remain on the exposed path, as shown in \figref{fig:overview-bwd}.

We model the MoE compute terms using the busiest post-reroute rank, with $T_{\text{moe}}^{bwd}\approx 2T_{\text{moe}}^{fwd}$ to account for both Wgrad and Dgrad computation.
\begin{equation}
    T_{\text{moe}}^{fwd/bwd}
    \propto
    \max_{r\in\mathcal{R}}
    \sum_{e\in\mathcal{E}} u_{e,r}.
\label{eq:tmoe}
\end{equation}

Similarly, token all-to-all is dominated by the busiest sender or receiver rank, and backward follows the same imbalance pattern as forward under the same load distribution.
\begin{equation}
    T_{\text{tok\_a2a}}^{fwd/bwd}
    \propto
    \max_{r\in\mathcal{R}}
    \max\!\left(
        \sum_{e\in\mathcal{E}} \lambda_{r,e},
        \sum_{e\in\mathcal{E}} u_{e,r}
    \right).
\label{eq:tcomm}
\end{equation}
In training, the number of tokens sent by each rank $\sum_e \lambda_{r,e}$ is fixed by the microbatch shape and parallelism configuration; in prefill, it is upper-bounded by the chunked-prefill size.

Since weight-distribution latency is dominated by the rank hosting the hottest main experts, we approximate its cost by
\begin{equation}
T_{w\_\text{distr}}^{fwd}
\propto
\max_{r\in\mathcal{R}}
\sum_{e \in \mathcal{E}_{r}} \left(|\mathcal{H}(e)| - 1\right),
\label{eq:twdistr}
\end{equation}

\parabf{Constraints.}
First, main expert placement is immutable. Second, each rank has a redundant-slot budget of $N_{\mathrm{slot}}$, and no logical expert may appear more than once on the same rank.
Third, the backward communication introduced by replicas must be fully hidden to ensure no visible overhead is created in backward. Fourth, we additionally require every newly created replica to carry at least a quota of $u_{\min}$.






\subsection{Practical Challenges}
\label{subsec:overview-challenges}

\paraf{Control Plane: Tiny Decision Window.}
Real-time balancing uses exact load and thus avoids prediction error.
However, solving the plan must deliver high balancing quality under a tight latency budget.
We resolve this challenge in \secref{sec:method-solver}.
{
\setlength{\leftmargini}{1.2em}
\begin{itemize}
    \item \uline{\emph{Combinatorial Decision Space at Large-EP.}}
        Prior systems decouple placement and reroute.
        Placement is planned ahead with sufficient budget, while online reroute either applies round-robin allocation or patches prediction drift under strict replica-count constraints~\cite{lplb}.
        With exact load, optimal balance requires replication to provision reroute capacity with flexible replica candidates, substantially enlarging the decision space at large-EP.
    \item \uline{\emph{Robustness to Non-Stationary Loads.}}
        Hot experts can shift quickly across microbatches.
        The solver must remain stable and fast under frequent distribution changes, rather than overfitting to transient patterns.
\end{itemize}
}

\parabf{Data Plane: RSN Bandwidth Underutilization.}
RSNs offer high scale-up bandwidth, yet balancing traffic is asymmetric and highly dynamic, exposing new system bottlenecks.
Existing communication stacks are optimized for static, predictable collectives and adapt poorly to the irregular traffic in \sys.
We resolve this challenge in \secref{sec:method-comm}.
{
\setlength{\leftmargini}{1.2em}
\begin{itemize}
    \item \uline{\emph{Magnified Non-Payload Overhead.}}
        For irregular weight or gradient synchronization on RSNs, the bottleneck shifts from link bandwidth to control-path overheads, including task tiling and pipelining, data staging, and dynamic addressing, preventing bandwidth saturation.
        Blindly scaling up parallelism is wasteful and can contend with overlapped backward computation for the same resources.

    \item \uline{\emph{Fan-Out Bottleneck of Hot Experts.}}
        Expert replication induces a dynamic subgroup multicast pattern.
        Under large-EP and skewed load, popular main experts with many replicas can incur large fan-out overhead that offsets balancing gains.
        Some RSNs support offloading multicast to the switch, but this typically assumes predetermined communication groups and delivery modes~\cite{nv-sharp}.
        In contrast, expert replication yields sparse and volatile groups that can change across layers and microbatches.
\end{itemize}
}

\section{Quota-Driven Planning}
\label{sec:method-solver}

\sys plans balancing as a joint replication-reroute problem driven by exact runtime load.
Instead of deciding replicas first and then patching them with a separate reroute policy, we directly solve the final per-instance load quota $U$.
Replication materializes the physical instances needed to realize $U$, and reroute then induces a rank-wise split $q$ consistent with the quota targets.
\algoref{algo:quota-solver-yushan} depicts the full picture.

\subsection{Replication}
\label{subsec:method-solver-rep}

\parabf{Threshold Formulation.}
Let $\lambda_e=\sum_{r\in\mathcal R}\lambda_{r,e}$ denote the total load of expert $e$, and $\ell_r=\sum_{e\in\mathcal{E}_{r}} \lambda_e$ the initial load on rank $r$.
We seek the smallest load threshold $\tau$ such that every rank can be brought below $\tau$ using replications alone.
For a candidate $\tau$, define
\begin{equation}
    \mathrm{exc}_r(\tau)=\max(\ell_r-\tau,0),\qquad
    \mathrm{slk}_r(\tau)=\max(\tau-\ell_r,0).
\label{eq:solver-excess-slack}
\end{equation}
Here $\mathrm{exc}_r(\tau)$ and $\mathrm{slk}_r(\tau)$ denote the excess load rank $r$ must shed and slack load it can still absorb, respectively.
A threshold $\tau$ is feasible if excess load can be reassigned from overloaded ranks to admissible ranks without violating slack, the per-rank slot budget, or the no-duplicate constraint, such that every rank ends with load at most $\tau$.

\parabf{Quota Construction.}
\sys binary-searches $\tau$ between the target rank load and the initial maximum rank load, and runs a greedy feasibility oracle for each probe.
For a candidate $\tau$, the oracle visits overloaded ranks by descending residual excess and their main experts by descending $\lambda_e$.
It transfers as much load as possible from the hottest remaining expert to the admissible rank with the largest slack, subject to quota floor $u_{\min}$.
Accepted transfers both create a replica and update its quota in the temporary plan $\tilde U$, so the placement already encodes useful reroute capacity.
If all residual excess is drained, the probe records $\tilde U$ and searches for a smaller threshold; otherwise the probe is infeasible and the search moves upward.
\uline{\emph{Why efficient:}}
By using quota as the coupling variable, each probe reserves reroute capacity while choosing replicas.
This avoids enumerating replica sets or token-level routes, and prevents ineffective replicas that would satisfy placement heuristics but receive little traffic.

\begin{algorithm}[t]
\caption{Replication\&Reroute Joint Solving}
\label{algo:quota-solver-yushan}
\footnotesize
\DontPrintSemicolon
\SetKwProg{Fn}{Function}{}{}
\SetKwFunction{SolveReplication}{SolveReplication}
\SetKwFunction{SolveReroute}{SolveReroute}
\SetAlgoNlRelativeSize{-1}
\KwIn{Load $\Lambda=\{\lambda_{r,e}\}$, $N_{\mathrm{slot}}$, quota floor $u_{\min}$, balancing target $\beta$}
\KwOut{Slot assignment $X$, quotas $U$, reroute split $Q$}

\Fn{\SolveReplication{$\Lambda, N_{\mathrm{slot}}, u_{\min}, \beta$}}{
    Precompute $\lambda_e\gets\sum_r \lambda_{r,e}$ and $\ell_r\gets\sum_{e\in\mathcal E_r}\lambda_e$\;
    $\tau_{\mathrm{lo}}\gets \beta \cdot \left\lceil \frac{1}{R}\sum_r \ell_r \right\rceil$, \quad $\tau_{\mathrm{hi}}\gets \max_r \ell_r$\;
    \While{$\tau_{\mathrm{lo}}<\tau_{\mathrm{hi}}$}{
        $\tau \gets \left\lfloor \frac{\tau_{\mathrm{lo}}+\tau_{\mathrm{hi}}}{2} \right\rfloor$\;
        $\mathrm{exc}_r\gets \max(\ell_r-\tau,0)$, $\mathrm{slk}_r\gets \max(\tau-\ell_r,0)$\;
        Initialize $\tilde u_{e,h(e)}\gets \lambda_e$ and $\tilde u_{e,t}\gets 0$ for $t\neq h(e)$\;
        \ForEach{overloaded rank $r$ in descending $\mathrm{exc}_r$}{
            \ForEach{main expert $e\in\mathcal{E}_r$ in descending $\lambda_e$}{
                $\mathrm{cap}_e\gets$ remaining transferable load of $e$\;
                \While{$\mathrm{exc}_r>0$ and $\mathrm{cap}_e>0$}{
                    $\mathcal{T}\gets$ admissible host ranks under slack, slot, and no-duplicate constraints\;
                    \lIf{$\mathcal{T}=\varnothing$}{\textbf{break}}
                    $t^\star\gets \arg\max_{t\in\mathcal{T}} \mathrm{slk}_{t}$\;
                    $\delta\gets \min(\mathrm{exc}_r,\mathrm{slk}_{t^\star},\mathrm{cap}_e)$\;
                    \lIf{$\delta<u_{\min}$}{\textbf{break}}
                    $\tilde u_{e,h(e)} \gets \tilde u_{e,h(e)}-\delta$\;
                    $\tilde u_{e,t^\star} \gets \tilde u_{e,t^\star}+\delta$\;
                    $\mathrm{exc}_r -= \delta$, $\mathrm{slk}_{t^\star} -= \delta$, $\mathrm{cap}_e -= \delta$\;
                }
            }
        }
        \eIf{$\sum_r \mathrm{exc}_r=0$}{
            $U \gets \tilde U$, $\tau_{\mathrm{hi}}\gets \tau$ \tcp*[r]{Feasible}
            Materialize $\tilde U$ into slot assignment $X$\;
        }{$\tau_{\mathrm{lo}}\gets \tau+1$ \tcp*{Infeasible}}
    }
    \Return{$X,U$}\;
}

\BlankLine

\Fn{\SolveReroute{$\Lambda,U$}}{
    \ForEach{logical expert $e$}{
        \ForEach{host rank $t \in \mathcal{H}(e)$}{
            $q_{t,e,t} \gets \min(\lambda_{t,e}, u_{e,t})$ \tcp*{Consume local quota}
            $\hat{\lambda}_{t,e} \gets \lambda_{t,e} - q_{t,e,t}$ \tcp*{Residual demand}
            $\hat{u}_{e,t} \gets u_{e,t} - q_{t,e,t}$ \tcp*{Residual quota}
        }
        \ForEach{source rank $r$ with $\hat{\lambda}_{r,e}>0$}{
            \ForEach{host rank $t \in \mathcal{H}(e)$ with $\hat{u}_{e,t}>0$}{
                $q_{r,e,t} \gets \text{round}\left( \hat{\lambda}_{r,e} \times \frac{\hat{u}_{e,t}}{\sum_{t'} \hat{u}_{e,t'}} \right)$\;
            }
        }
    }
    Materialize $Q=\{q_{r,e,t}\}$ with per-token assignment.\;
    \Return{$Q$}\;
}

\BlankLine
$(X,U)\gets \SolveReplication(\Lambda,N_{\mathrm{slot}},u_{\min},\beta)$\;
$Q\gets \SolveReroute(\Lambda,U)$\;

\Return{$X,U,Q$}\;

\end{algorithm}

\subsection{Reroute}
\label{subsec:method-solver-reroute}

\parabf{Quota Decomposition with Locality.}
Once $U$ is fixed, reroute no longer revisits the balancing objective; it only materializes a source-wise split $Q$ whose aggregate load matches the solved quotas.
For each expert, \sys first lets tokens originating from the same host rank consume that host's quota.
Prioritizing local quota only changes which source rank consumes a solved quota, not the quota itself.
Therefore, token locality reduces cross-rank traffic without breaking the solved threshold.
Reroute then distributes the residual source demand over the remaining quotas in proportion to their residual capacity, with deterministic rounding to preserve both per-source demand and per-instance quota.

\parabf{Token Assignment.}
Quotas specify aggregate source-to-instance counts, whereas all-to-all dispatch requires a per-token destination.
Each rank stores the resulting decomposition as cumulative quotas ordered by the physical instances of each logical expert with a lightweight prefix scan.
During dispatch, the $j$-th local token of pair $(r,e)$ is sent to the first physical instance whose cumulative quota covers $j$.
This reduces token assignment to a small, rank-localized upper-bound lookup, independent of the optimization procedures.

\subsection{GPU-Native Solving}

\sys implements quota solving fully on device, avoiding CPU synchronization and device-host metadata movement on the hot path.
The main challenge is that the algorithm is not a simple data-parallel scan: each binary-search probe mutates excess, slack, slot occupancy, and per-expert replica sets, while the feasibility oracle has sequential dependencies across accepted transfers.
\sys exploits warp-level parallelism and data locality. With one cooperative thread block on a single streaming multiprocessor (SM), it stages the load matrix and placement state in shared memory, evaluates multiple threshold probes across warps, and uses warp-level reductions to find admissible high-slack targets under the slot and no-duplicate constraints.
The same kernel emits the final slot mapping, quotas, and cumulative reroute metadata, turning quota solving from a CPU-side combinatorial search into a compact GPU-resident feasibility problem.

\section{RSN-Native Balancing Communication}
\label{sec:method-comm}

\sys's balancing traffic is a runtime-adaptive sparse transfer graph over RSN scale-up fabrics, rather than a regular collective.
It serves two paths: (1) both forward weight distribution and backward redistribution from each main expert to remote replicas, and (2) backward gradient reduction from those replicas back to main experts.
The main goal is therefore high \emph{effective} bandwidth under a volatile per-layer plan, not merely peak link rate.
We optimize the two bottlenecks identified in \secref{subsec:overview-challenges}: non-payload overhead (\secref{subsec:method-comm-pipeline}) and hot-expert fan-out (\secref{subsec:method-comm-relay}).

\subsection{Persistent Tile Streaming for Data Transfers}
\label{subsec:method-comm-pipeline}

Both weight distribution and gradient reduction use the same execution unit: the weight or gradient of each expert is divided into fixed-size \emph{tiles}, and the solved placement plan is compiled into device-resident transfer tasks over these tiles.
Instead of issuing a transfer per replica, \sys runs a persistent kernel whose thread blocks repeatedly pull the next tile from the global task stream.
For weight distribution, each source tile is staged once into shared memory and stored to all remote replica destinations.
For gradient reduction, gradient tiles from remote replicas are loaded and accumulated into the local gradient buffer of main experts, and then cleared in place for reuse in later layers.
The kernels double-buffer the shared-memory tile within each thread block: while tile $i$ is being stored or reduced, the block fetches the next tile index and starts the load for tile $i{+}1$.
Thus, plan-dependent task lookup, address translation, and synchronization overhead are folded into the tile pipeline and hidden by data movement, rather than exposed as separate control steps for every replica transfer.

\parabf{Overlap-Aware Footprint.}
The tile-streaming kernels expose the number of resident thread blocks as the main resource knob, which is sized with the overlap window.
On the forward critical path, weight distribution launches sufficient thread blocks to increase in-flight tile loads and remote stores.
On overlapped backward paths, the footprint is bounded explicitly with configurable SM residency.
The shared memory usage is also bounded to the active pipeline buffers, to minimize the contention with other compute-intensive backward kernels scheduled on the same SM.
Thus \sys can spend more occupancy on the hot path to saturate RSN scale-up bandwidth, while preserving enough headroom for concurrent overlapped kernels during backward.

\subsection{Chunk Streaming Relay for Hotspot Fan-Out}
\label{subsec:method-comm-relay}

\begin{figure}[t]
    \centering
    \includegraphics[width=0.98\linewidth]{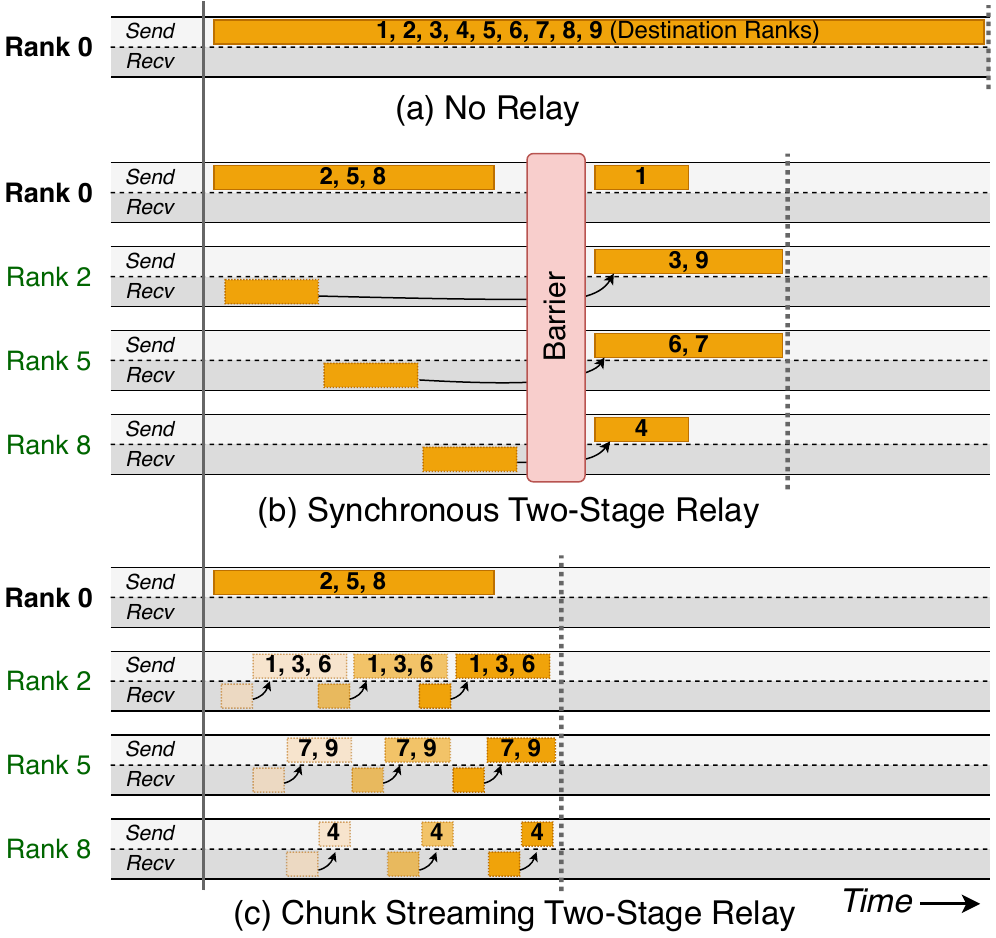}
    \caption{Relay schemes for hot expert fan-out, supposing one expert on rank~0 that multicasts to replica ranks~1--9, with ranks~2, 5, and~8 selected as relays.
    For clarity, the figure omits leaf ranks and finer-grained tiles.
    Each rank displays the state of send/receive channels along the timeline.}
    \Description{Chunk-streaming relay schedule for distributing a hot expert from a source rank to relay ranks and replica ranks.}
    \label{fig:tech-relay}
\end{figure}

Persistent tile streaming keeps each single transfer efficient, but a hot expert can still bottleneck on fan-out traffic.
The exposed bound is sender-side: each rank has at most $N_{\text{slot}}$ inbound replicas, but may need to push several hot experts to many destinations.
For experts whose replica count exceeds the relay threshold (set to 4), \sys builds a lightweight two-stage relay: the source rank first seeds a small relay set, and each relay then forwards to its assigned leaves.
The relay frontier is chosen near $\sqrt{|\mathcal{H}(e)|-1}$, which approximately balances the two stages and reduces the source's critical fan-out accordingly.
The relay scheduling works on \emph{chunks} of consecutive tiles rather than whole experts, so a relay rank can forward a chunk immediately after it arrives.
After all tiles of a stage-I chunk reach the relay rank, the kernel writes the per-chunk ready flag; stage-II waits on the corresponding flag and immediately sends that chunk from the local relay buffer to its leaves.
As shown in \figref{fig:tech-relay}, this chunk-level streaming pipelines the two stages without waiting for the whole expert or introducing global inter-stage barriers.

\parabf{Load-Aware Relay Scheduling.}
The kernel builds the relay topology that balances outgoing traffic of all ranks.
It first tracks the volume of sending bytes assigned to each rank by the replication plan, then processes relay-eligible hot experts one by one.
For each, it selects first-stage relays from the expert's replica ranks with the smallest sending volume.
The remaining replicas are attached to the relays whose sending volume would remain smallest after taking the new leaf.
Before moving to the next hot expert, the sending volumes of the source rank and its relays are updated to the send for relays and leaves, respectively. 
This scheduling only decides the relay trees, and each edge still adopts the chunk streaming transfer.
In this way, the hot expert's fan-out is split across ranks with spare sending capacity.

\section{Implementation}
\label{sec:impl}

We design \sys as a standalone runtime decoupled from both training/serving frameworks and MoE token all-to-all backends.
The core library contains about 9.6K lines of C++ (including device kernels) and Python code.
We integrate \sys into Megatron-LM~\cite{Shoeybi2019Megatron} for training and SGLang~\cite{zheng2024sglang} for serving; both stay below 1K lines of additional code.
We use DeepEP~\cite{deepep} (\texttt{hybrid-ep} branch optimized for intra-rack communication, v1.2.1+7febc6e) for token dispatch/combine in both frameworks.
Being fully on-device, \sys avoids host-device transfers and preserves the graph capture of device operations.

\parabf{RSN Memory Semantics.}
For communication over RSN scale-up fabrics, we adopt GPU-initialized, one-sided peer-memory access.
At initialization, all ranks allocate symmetric buffers for redundant weights/gradients, placement/load metadata, and other flags, then resolve intra-RSN peer handles into device-resident address tables.
Transfer kernels consume compact task descriptors and access peer buffers through load/store primitives.

\parabf{End-to-End Integration.}
\sys maintains redundant experts as layer-shared internal buffers, while persistent model states of main experts are backed by external frameworks.
These redundant experts are excluded from framework-side parameter/gradient buckets, optimizer state, and checkpoints.
After weight initialization in SGLang or bucket construction in Megatron-LM, \sys lazily registers the weight and gradient pointers of main experts during model forward.
For backward, \sys dynamically assigns each in-flight MoE invocation a \emph{virtual layer ID}, which hashes the placement and reroute metadata with the specific (real layer, microbatch) in a ring buffer.
Carried through \texttt{torch.autograd}, this ID allows backward weight re-materialization and gradient reduction to retrieve the matching forward balancing plan from \sys's internal state.
By setting the ring size to the maximum in-flight microbatches, we accommodate PP and virtual PP~\cite{narayanan2021efficient} while keeping \sys agnostic to PP scheduling details across microbatches and stages.
Because \sys only operates within EP groups, it remains orthogonal to attention-side DP, TP, and model-wide DP.

%% file: sections/5-evaluation.tex
\section{Evaluation}
\label{sec:evaluation}

\input{sections/table-setup}

\begin{figure*}[t]
    \centering
    \includegraphics[width=0.88\linewidth]{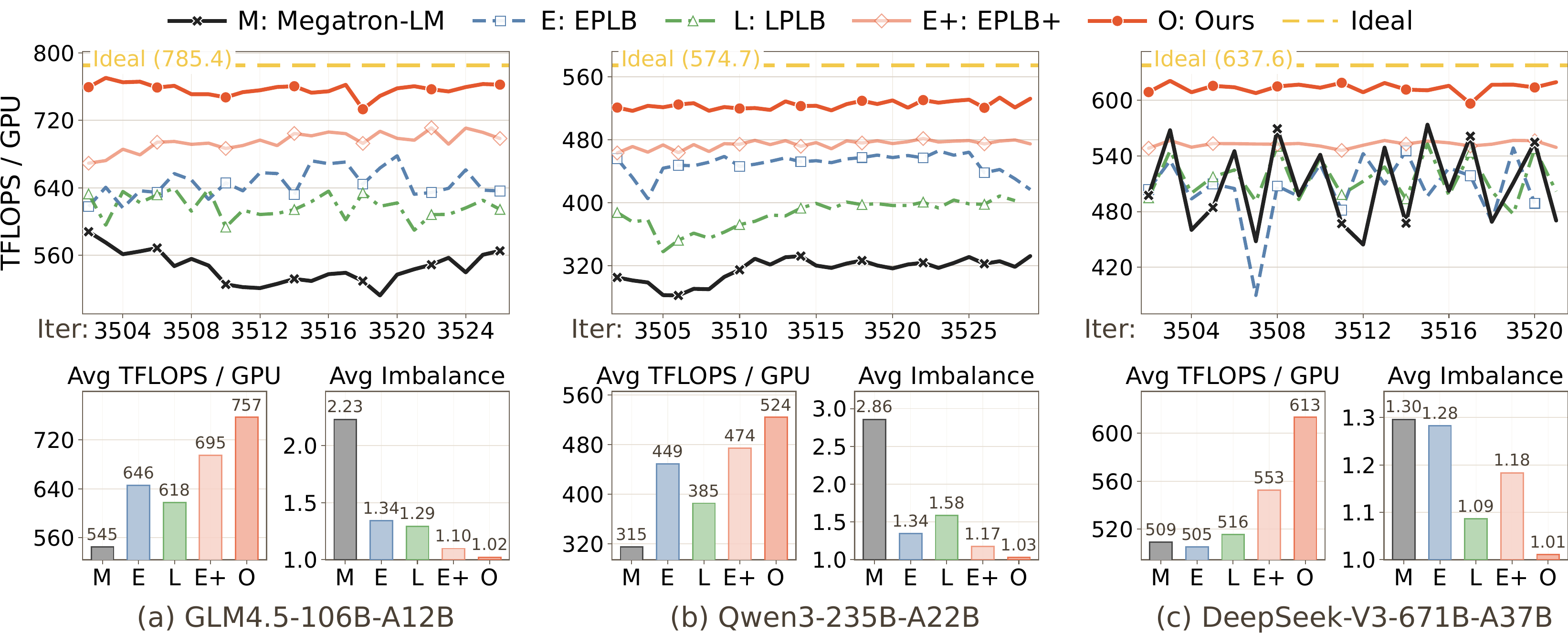}
    \caption{\textbf{End-to-End Training Performance:} Varying throughput across 20 training iterations on three models.}
    \Description{Late-stage training throughput across models and methods.}
    \label{fig:eval-e2e-train}
\end{figure*}

\begin{figure*}[t]
    \centering
    \includegraphics[width=0.95\textwidth]{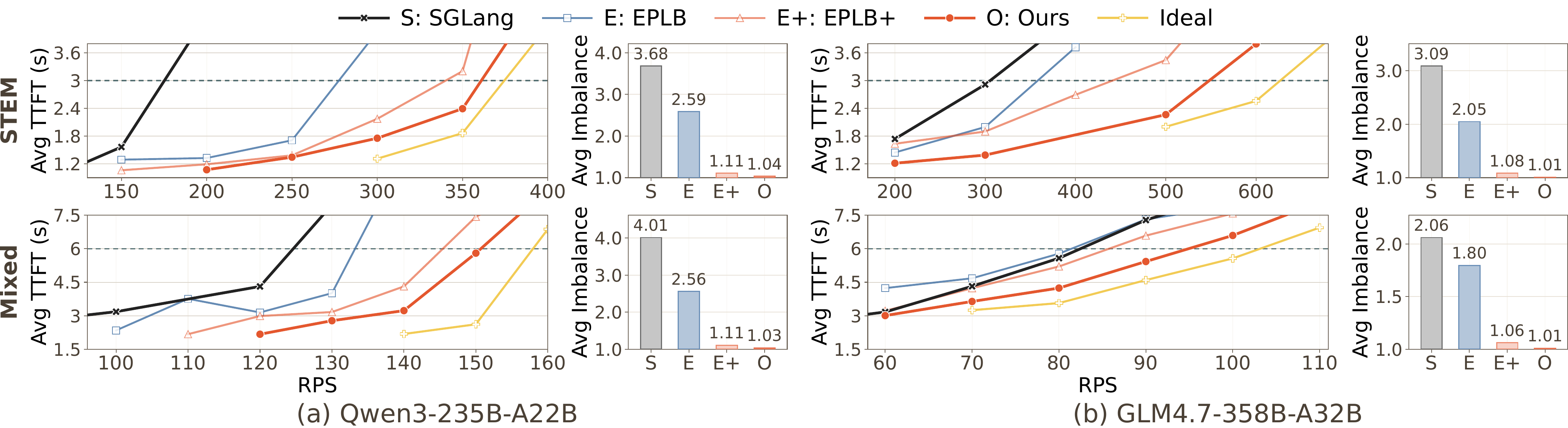}
    \caption{\textbf{End-to-End Prefill Performance:} RPS--mean TTFT trade-offs on two data domains and two models.}
    \Description{RPS and mean TTFT trade-offs for serving prefill across models, data domains, and methods.}
    \label{fig:eval-e2e-serving}
\end{figure*}

\subsection{Setup}
\label{subsec:eval-setup}

\parabf{Testbed.}
Our evaluation runs on a public-cloud RSN cluster, where each rack contains 64 GPUs (16 servers).
The bandwidth of scale-up links is 8--10 times of the scale-out RDMA network.
For research prototyping, serving prefill uses one rack, whereas training uses two or four racks.
We also perform production-scale MoE training spanning multiple racks.

\parabf{Models and Parallelism.}
We evaluate four MoE models spanning diverse scales, sparsity levels, and parallelism configurations.
\tabref{tab:eval-models} summarizes their key characteristics.
All experiments use bf16 precision.
For \uline{serving}, we evaluate Qwen3-235B~\cite{yang2025qwen3} and GLM4.7-358B~\cite{glm2026v47}.
Each runs a prefill server within one rack, under EP64 or EP40 and DP attention.
For \uline{training}, we benchmark GLM4.5-106B~\cite{glm2025v45} on 128 GPUs (2 racks), and Qwen3-235B~\cite{yang2025qwen3} plus DeepSeek-V3~\cite{deepseek2024v3} on 256 GPUs (4 racks).
In production, we train an in-house model called RefMoE-288B-A16B under EP32.
Training uses intra-rack EP, with DP- or PP-based inter-rack scaling.

\parabf{Training Recipes.}
For prototyping, we train the three open-source models with Megatron-LM, disabling system-side balancing while saving intermediate checkpoints for evaluation.
Each run uses a 200B-token subset from our in-house corpus, and lasts about 4500 global batches, with the batch size ramping from 1024 to 5120.
In production, we train RefMoE-288B with our internal training stack and enable \sys.
GLM4.5-106B and Qwen3-235B use the GShard load-balancing loss with weight $10^{-2}$, whereas DeepSeek-V3 and RefMoE follow the DeepSeek recipe with routing bias update speed $10^{-3}$ and sequence-level loss weight $10^{-4}$.

\parabf{Serving Workloads.}
We construct queries from realistic reasoning workloads:
(1) \uline{STEM}, including coding (Codeforces~\cite{codeforces}, SWE-bench~\cite{jimenez2024swebench}), mathematics (DAPO-Math-17K \cite{yu2025dapo}), science (GPQA~\cite{rein2024gpqa}, OpenScience~\cite{nvidia-openscience}), and (2) \uline{Mixed}, with additional multi-task, long-context queries from LongBench~\cite{bai2024longbench}.
Input lengths range from several hundred to tens of thousands of tokens.
We generate traces with a Poisson arrival process at different request rates.

\parabf{Baselines.}
We compare \sys against the following baselines, with tuned compatibility on the underlying RSNs.

{
\setlength{\leftmargini}{1.2em}
\begin{itemize}
    \item \textbf{Megatron-LM}~\cite{Shoeybi2019Megatron}: \texttt{dev} branch, commit \texttt{e93814b}.
    \item \textbf{SGLang}~\cite{zheng2024sglang}: \texttt{main} branch, \texttt{v0.5.9+bbe9c7e}.
    \item \textbf{EPLB}~\cite{eplb}: a widely used algorithm for computing balanced expert placement plans, based on recent load.
    We optimize its integration into SGLang and Megatron-LM on RSNs for negligible balancing overhead.
    We use 50 prefill steps and 3 global batches as the rebalancing frequency for serving and training, respectively.
    \item \textbf{LPLB}~\cite{lplb}: a linear-programming solver that augments EPLB with reroute adjustment for each microbatch in training.
    It enforces at most one replica per expert to reduce overhead.
    We pair it with EPLB in Megatron-LM.
    \item \textbf{EPLB$+$} (w/ exact load): to isolate the benefit of our exact-load quota solving, we replace \sys's planning with standard EPLB and round-robin reroute while keeping the communication mechanism unchanged.
    \item \textbf{Ideal}: a force-balanced upper bound in Megatron-LM and SGLang, where we modify the router to dispatch tokens evenly across experts for perfect balancing.
\end{itemize}
}

\parabf{Metrics.}
For \uline{training}, we report achieved throughput in TFLOPS/GPU by resuming late-stage checkpoints and running a short continuation window representative of the whole training process.
This avoids resource-intensive full training of every balancing baseline.
For \uline{serving}, we report overall prefill latency, i.e., TTFT as a function of requests per second (RPS).
We also report per-rank max/mean imbalance ratios, averaged across all layers and batches.

\subsection{End-to-End Performance}
\label{subsec:eval-e2e}

Across training and serving prefill, \sys sustains an average 94.6\,\% and 93.9\,\% of the ideal performance and consistently outperforms all baselines under varying load.

\parabf{Training (\figref{fig:eval-e2e-train}).}
To evaluate steady-state training performance, each baseline resumes the same model checkpoint at the 3500$^{\mathrm{th}}$ global batch and runs for 20 more, which is long enough to cover multiple EPLB balancing intervals and capture the dynamic load changes.
On all three models, \sys keeps the average per-rank imbalance at only 1.01--1.03 after balancing and delivers stable throughput across global batches.
In contrast, baseline throughput oscillates visibly because their balancing plans lag the realized hot experts and the resulting stragglers change over time.
Averaged across three models, EPLB, LPLB, EPLB$+$, and \sys improve throughput by 20\,\%, 12\,\%, 29\,\%, and 42\,\% over Megatron-LM, respectively.
LPLB is constrained by its limited replica budget and solving overhead.
For DeepSeek-V3 (c), routing compensation reduces the overall imbalance but enlarges short-term load swings (\secref{sec:load-profile}), where EPLB and LPLB show similar or even worse performance than Megatron-LM, while \sys remains above 96\,\% of the ideal.
The remaining gap to force-balancing mainly comes from uneven routing in realistic MoE training, instead of residual imbalance or hot-path balancing overhead; we further analyze this in \secref{subsec:eval-latency}.

\parabf{Serving Prefill (\figref{fig:eval-e2e-serving}).}
Prefill load is even more skewed and non-stationary than training because request semantic domains, prompt lengths, batch composition, and arrival times all change drastically.
Under this stronger turbulence, \sys yields larger gains, and still reaches 90\,\%--97\,\% of ideal throughput.
Compared with SGLang and EPLB, \sys achieves 1.56$\times$ and 1.29$\times$ higher throughput, respectively.
\sys also consistently outperforms EPLB$+$, with 5\,\%--24\,\% gains.
To ensure identical load conditions in balancing quality comparisons, we record the full routing trace of SGLang under full load, and replay it under other evaluated balancing algorithms.
\sys sustains 1.01--1.04 realized imbalance across all models and domains.

\subsection{Latency Breakdown}
\label{subsec:eval-latency}

\begin{figure}[t]
    \centering
    \includegraphics[width=0.95\linewidth]{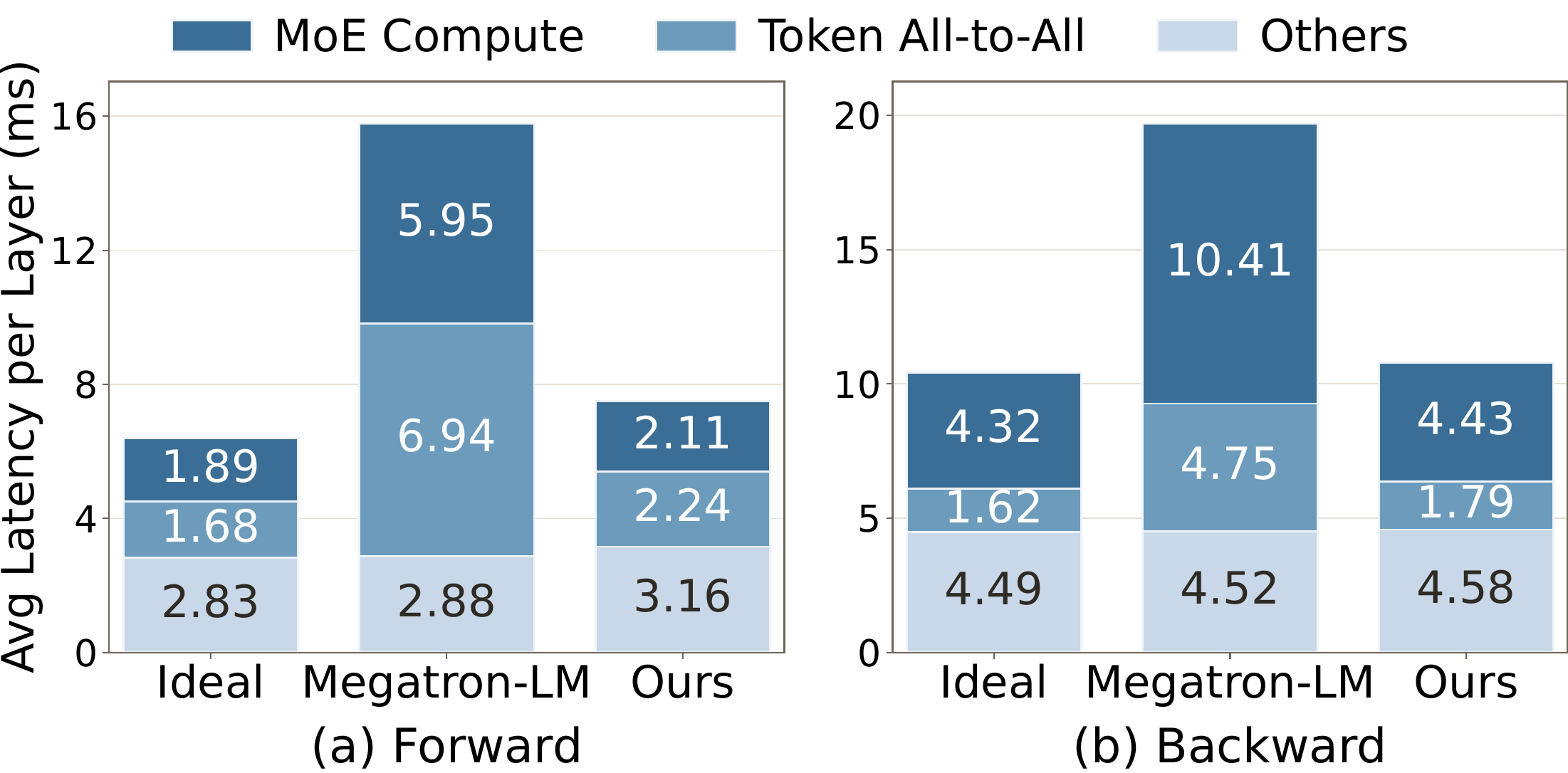}
    \caption{Latency breakdown of forward and backward passes during Qwen3-235B-A22B training.}
    \Description{Stacked time breakdown of attention, MoE communication, and MoE compute.}
    \label{fig:eval-time-breakdown}
\end{figure}

In \figref{fig:eval-time-breakdown}, we decompose the forward and backward latency of Qwen3-235B-A22B training, averaged per MoE layer, to illustrate the impact of \sys in depth.
Without balancing, Megatron-LM suffers large inflation in both MoE compute and token all-to-all.
With \sys, we first inspect the non-MoE part.
Compared with the ideal, the extra latency is 0.33 ms in forward and negligible in backward, which constitutes only 1.8\,\% of the total latency.
This indicates minimized hot-path overhead and the effectiveness of overlapping.
The MoE compute term is already close to the ideal, with slight control overhead induced by redundant physical experts.
This confirms that rank-level load imbalance is mostly removed.
For the remaining token all-to-all term, \sys increases latency by 33\,\% and 10\,\% in forward and backward, respectively.
This stems from uneven token routing in reality, distinct from synthetic uniform dispatch in the ideal.
For DeepEP, this irregularity translates into minor stalls within its internal token dispatch pipelines.
Overall, \sys eliminates most load imbalance with negligible hot-path overhead and moderate increase in compute and communication latency.

\subsection{Activation Memory Footprint}
\label{subsec:eval-memory}

\begin{figure}[t]
    \centering
    \includegraphics[width=0.95\linewidth]{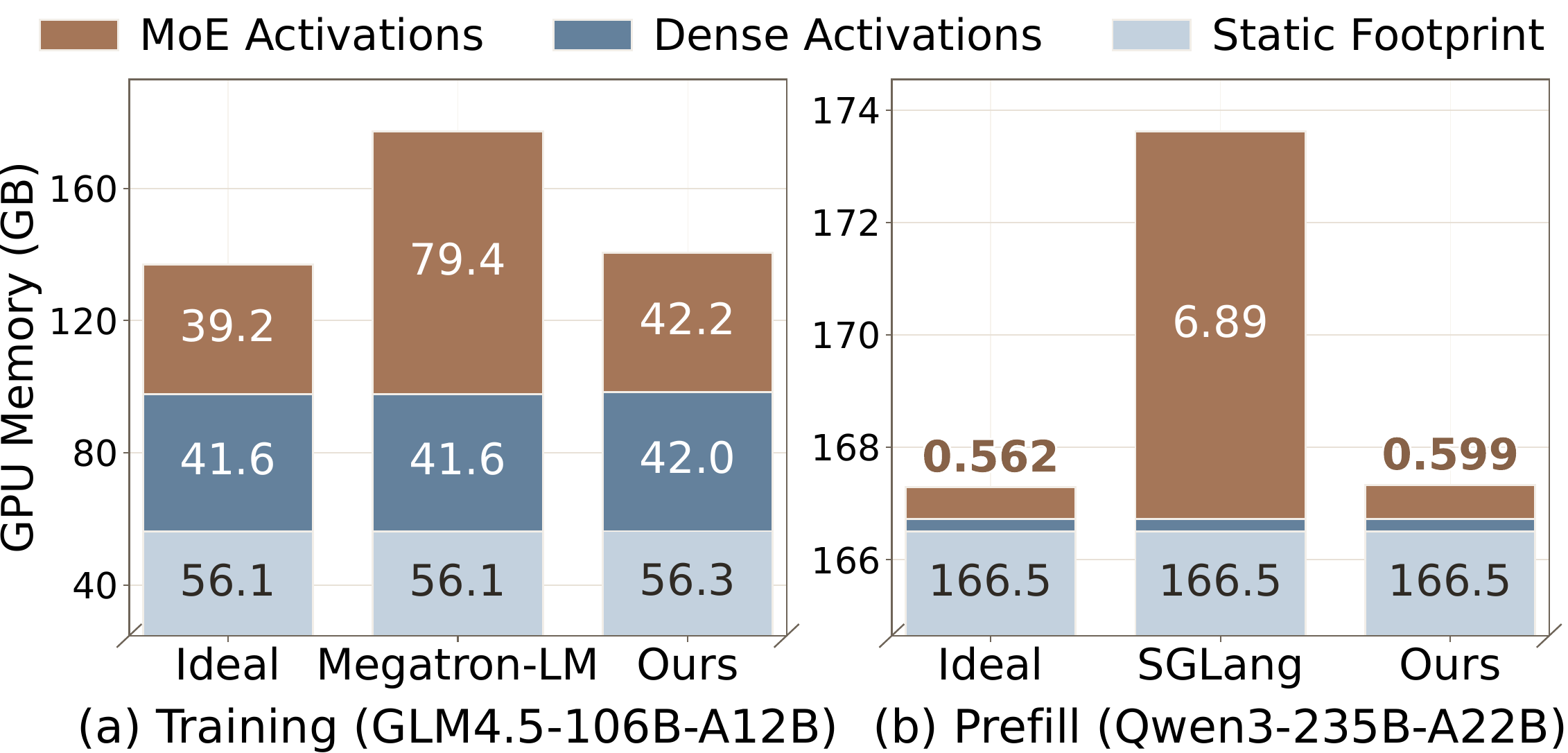}
    \caption{Breakdown of peak GPU memory.}
    \Description{Memory breakdown highlighting MoE receive-token activations.}
    \label{fig:eval-memory-saving}
\end{figure}

Besides latency, balancing also reshapes the peak activation footprint, especially on the hottest receiving ranks.
\figref{fig:eval-memory-saving} breaks down the peak GPU memory in both training and prefill for two models and highlights the MoE-related activation component.
For training, we disable activation checkpointing\footnote{Activation checkpointing trades memory savings with extra compute: it discards some forward activations and recomputes them in backward~\cite{chen2016sublinear}.} to expose the layer-accumulated activation upper bound, and the witnessed peak is during initial stages with high load skewness.
In contrast, serving only keeps transient activation of the current layer during forward.
Without balancing, we observe 2$\times$ and 11$\times$ higher peak memory of MoE activation than the ideal for training and serving, respectively.
By flattening receive-side hot spots, \sys substantially reduces the MoE activation peak and remains close to the ideal.
This directly lowers out-of-memory risk, improves model scaling headroom, and can avoid the extra performance loss from activation checkpointing.

\subsection{Ablation Study}
\label{subsec:eval-ablation}

\begin{figure}[t]
    \centering
    \includegraphics[width=0.96\linewidth]{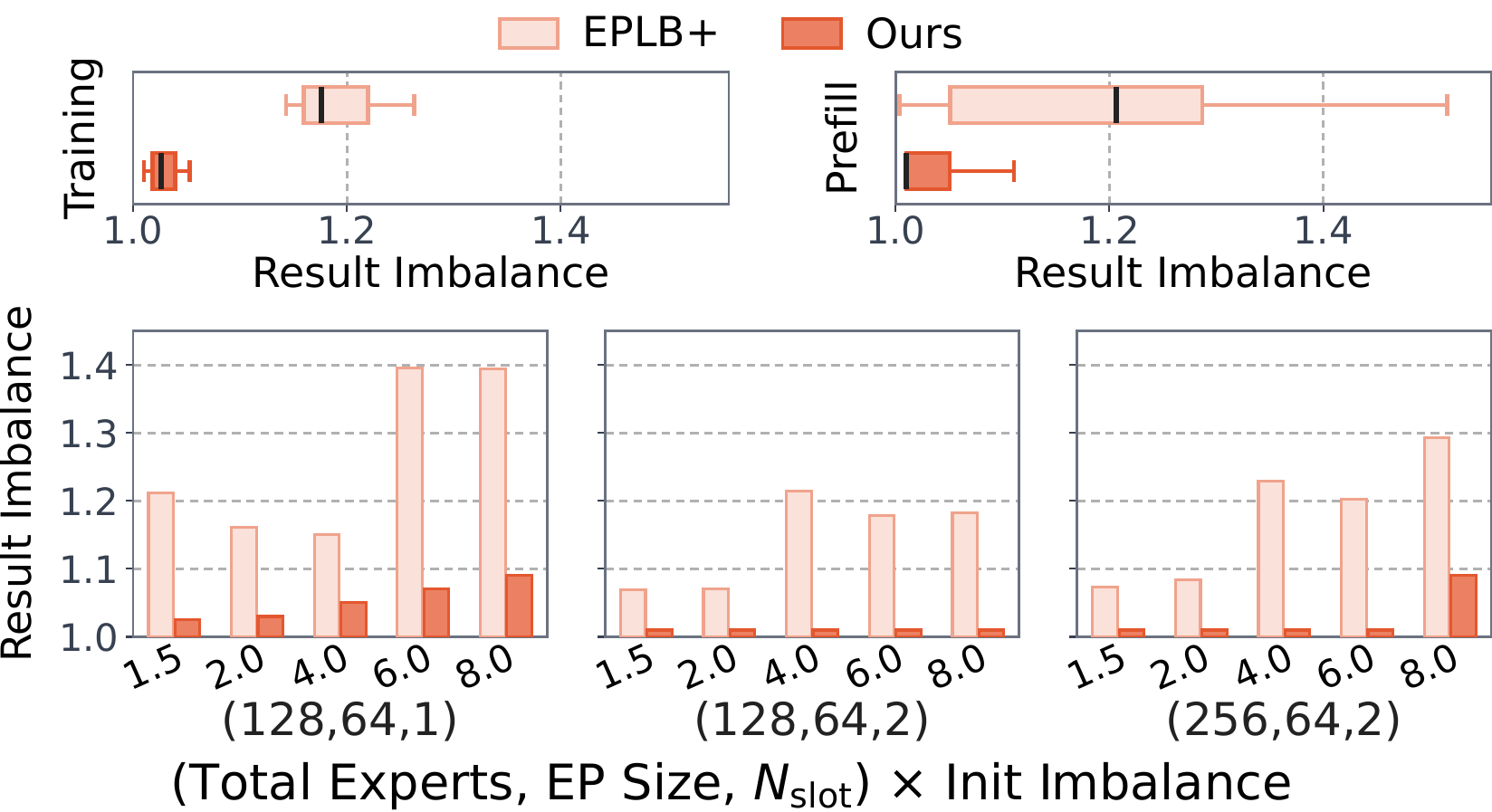}
    \caption{In-depth comparison of balancing quality between EPLB$+$ and \sys. \uline{Upper}: imbalance distribution of all training and prefill evaluations; \uline{Lower}: simulated balancing effect across various MoE, EP, and redundancy settings.
    The initial load is synthesized with a power-law distribution that resembles realistic skewness in MoE routing.}
    \Description{Balancing quality comparison between EPLB plus and UltraEP across measured and simulated settings.}
    \label{fig:eval-ablation-solver}
\end{figure}

\begin{table}[t]
    \caption{Balancing metrics averaged across simulations in \figref{fig:eval-ablation-solver}, including solving time, consumed redundant slots, maximum replica fan-out, and traffic ratio (which denotes effective in-flight tokens not absorbed by local ranks).}
    \label{tab:eval-ablation-solver}
{\small
\begin{tabular}{@{}lcc@{}}
\toprule
\textbf{Metrics (Avg)}       & \textbf{EPLB$+$}  & \textbf{Ours}                                                     \\ \midrule
Result Imbalance             & 1.19     & 1.03                                                                       \\
Solving Time (ms)            & 0.153    & 0.111                                                                      \\
$\sum_{e}{|\mathcal{H}(e)}|$ & 107      & 45                                                                         \\
$\max_{e}{|\mathcal{H}(e)}|$ & 8.5      & 6.8                                                                        \\
In-flight Token Ratio        & 99.9\,\% & 96.0\,\% (98.4\,\% w/o locality) \\ \bottomrule
\end{tabular}
}
\end{table}

\parabf{Balancing Quality.}
We compare our quota solver with EPLB$+$ to isolate its benefits.
\figref{fig:eval-ablation-solver} shows comprehensive balancing quality, while \tabref{tab:eval-ablation-solver} summarizes other relevant metrics.
Across all training and prefill evaluations, \sys demonstrates a distribution closer to the ideal and a much smaller tail than EPLB$+$.
Under severe initial imbalance and tighter replica budget, EPLB$+$ shows much higher imbalance up to 1.4, while \sys still keeps it below 1.1.
At the same time, \sys saves solving latency by 27.4\,\%, consumes 57.9\,\% fewer redundant slots, and reduces token traffic by 3.9\,\% with locality.
This comes from \sys's direct optimization of the post-reroute load bound, which is the actual objective of balancing, rather than the pre-reroute imbalance that EPLB$+$ focuses on.
Unlike EPLB$+$, which blindly replicates experts based on pre-reroute hotness, \sys only materializes a replica when it brings sufficient balancing gain.
This accounts for \sys's resource efficiency, which significantly reduces expert replication traffic.

\begin{figure}[t]
    \centering
    \includegraphics[width=0.95\linewidth]{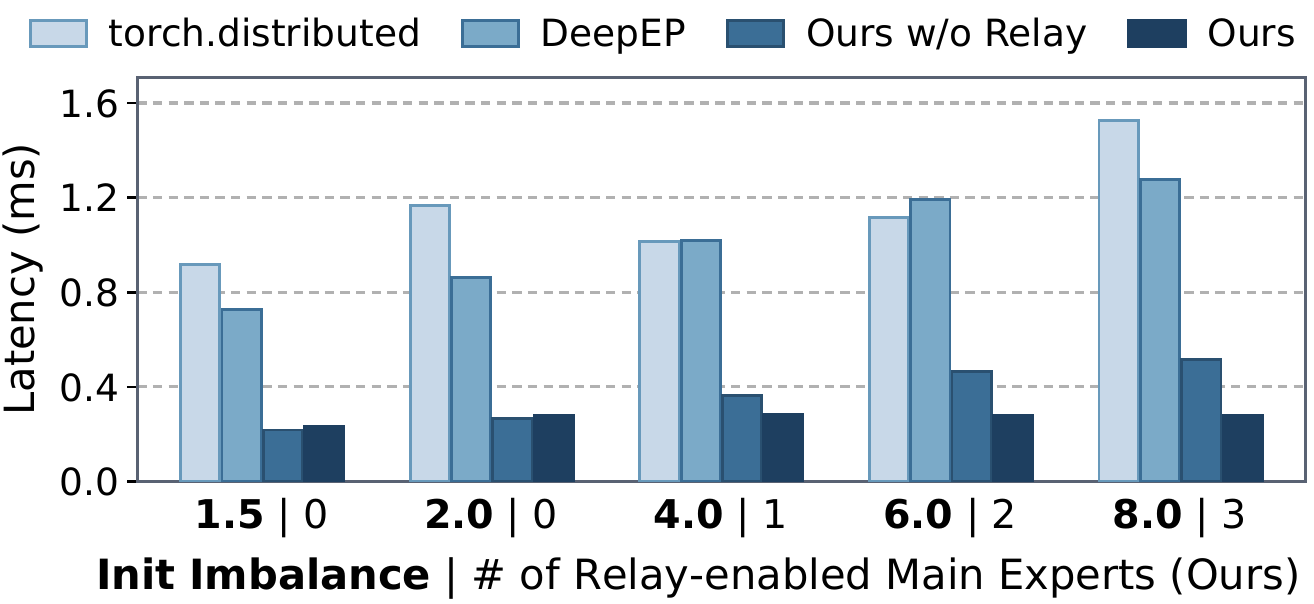}
    \caption{Communication latency of expert-weight distribution under various imbalance levels (as in \figref{fig:eval-ablation-solver}), comparing PyTorch distributed~\cite{li2020pytorch} batch send/recv, DeepEP, the no-relay ablation, and \sys, on Qwen3-235B and EP64.}
    \Description{Communication latency comparison for expert-weight distribution under different imbalance levels.}
    \label{fig:eval-ablation-comm}
\end{figure}

\parabf{Communication Performance.}
To isolate the effectiveness of our purpose-built, RSN-oriented communication optimizations, we adapt mainstream communication backends with RSN support for expert replication and compare them with \sys under identical balancing plans.
We tune DeepEP for expert transfer on top of its original token-dispatch substrate, and use PyTorch distributed~\cite{li2020pytorch} batch send/recv as a more general baseline.
As shown in \figref{fig:eval-ablation-comm}, \sys shows 3.1$\times$--5.5$\times$ speedup over \texttt{torch.distributed} and DeepEP under all imbalance levels.
For large-fan-out experts under high imbalance, enabling relay further provides 1.3$\times$--1.8$\times$ gains.
As the fan-out degree grows, \sys sustains near-constant latency around 0.28\,ms, while the no-relay variant shows linearly increasing latency.
Under lower imbalance, the adaptive policy keeps relay inactive and incurs only negligible control-path overhead for relay logic.
The relay design effectively mitigates the fan-out bottleneck.

\subsection{Production MoE Training}
\label{subsec:eval-prod}

\begin{figure}[t]
    \centering
    \includegraphics[width=0.95\linewidth]{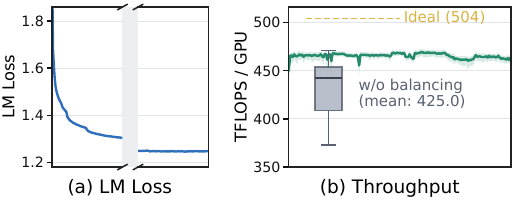}
    \caption{Throughput and loss over RefMoE-288B training process with \sys enabled.
    Panel (b) plots the stable training phase after batch-size ramp-up.
    We sample no-balancing throughput from continuation-run intervals with \sys disabled.
    For the ideal, we report the best measured force-balanced throughput to factor out environmental variability.
    }
    \Description{Throughput and loss curves for production training with UltraEP enabled.}
    \label{fig:eval-prod}
\end{figure}

We evaluate \sys's robustness and scalability in the real-world MoE training at larger scale than prototyping runs.
As shown in \figref{fig:eval-prod}, despite the hardware and network variability in the long run, \sys sustains over 92\,\% of the ideal throughput, with enhanced stability and 9.6\,\% average gain over no-balancing.
The loss curve follows the expected pretraining trajectory because \sys changes only the physical execution logic while preserving training semantics.
These results demonstrate that \sys can scale smoothly to production MoE training on multiple racks, while maintaining near-optimal load balancing and convergence.

%% file: sections/table-setup.tex
\begin{table}[t]
\centering
\caption{Evaluated models and parallelism settings.}
\label{tab:eval-models}
{\small
\setlength{\tabcolsep}{3.5pt}
\begin{tabular}{@{}lccc@{}}
\toprule
Model & \makecell{Experts\\(Top-$k$)} & \makecell{Parallelism\\(Train | Serve)} & $N_{\text{slot}}$ \\
\midrule
\textbf{GLM4.5-106B-A12B}~\cite{glm2025v45} & 128 (8) & \multicolumn{1}{l}{EP64-DP2 | --} & 2 \\
\textbf{Qwen3-235B-A22B}~\cite{yang2025qwen3} & 128 (8) & \multicolumn{1}{r}{EP64-DP4 | EP64} & 2 \\
\textbf{GLM4.7-358B-A32B}~\cite{glm2026v47} & 160 (8) & \multicolumn{1}{r}{-- | EP40} & 4 \\
\textbf{DeepSeek-V3-671B-A37B}~\cite{deepseek2024v3} & 256 (8) & \multicolumn{1}{l}{EP64-PP4 | --} & 2 \\
\bottomrule
\end{tabular}
}
\end{table}

%% file: sections/6-epilogue.tex
\section{Related Work}

\parabf{System-Side MoE Load Balancing.}
EPLB~\cite{eplb} and LPLB~\cite{lplb} are standalone balancers agnostic to load acquisition, computing expert layouts or reroutes with given load.
They are widely deployed, e.g., EPLB is fully supported in SGLang~\cite{zheng2024sglang} and vLLM~\cite{kwon2023vllm} for MoE EP serving.
Most research systems instead focus on providing such load estimates from history, cross-layer correlation, or profiled execution, then prefetch or re-layout expert shards before exact load is known~\cite{zhu2026probe,yang2026libra,zhang2025popfetcher,liu2026laer,nie2023flexmoe,zhai2023smartmoe,liu2023janus,he2022fastermoe}.
These predictive designs lack practicality for highly dynamic fine-grained MoE models at large-EP and RSN settings, whereas \sys reacts to realized load to achieve near-optimal balancing.

\parabf{MoE Computation and Communication Optimization.}
MoE communication libraries~\cite{deepep,mooncake-ep,mao2026ucclep,li2026swiftep} focus on specialized token all-to-all kernels across heterogeneous GPU and NIC platforms.
Computation optimizations mainly utilize kernel fusion to improve grouped-GEMM efficiency, or further fuse FFN execution with token dispatch and combine~\cite{gale2023megablocks,deepgemm}.
Other systems overlap MoE compute and token communication with fine-grained scheduling~\cite{he2022fastermoe,shi2023pipemoe,shi2024schemoe,pan2025fsmoe,jin2025megascale,deepseek2024v3,hwang2023tutel,zhang2025comet}.
These optimizations are orthogonal to and can be stacked with \sys for cumulative gains.

\section{Conclusion}
\sys enables exact-load expert balancing for large-EP MoE training and serving prefill on RSNs.
By coupling quota-based planning with RSN-native expert-state communication, \sys reacts to realized routing and performs real-time balancing, while keeping hot-path overhead small.
Our evaluation validates its near-ideal throughput, near-optimal balancing quality, and production scalability.
Because \sys covers both training and inference, the same abstraction can naturally extend to reinforcement learning (RL) pipelines that alternate these two procedures.